\documentclass[twocolumn,letterpaper,aps,prc,superscriptaddress,showpacs,nofootinbib,floatfix]{revtex4}

\usepackage{amsmath}
\usepackage{amssymb}

\usepackage{graphicx}
\usepackage{dcolumn}
\usepackage{bm}
\usepackage{verbatim}
\usepackage{lineno}

\newcommand{\pt}{$p_T$ }

\begin{document}


\title{Azimuthal correlations of electrons from heavy-flavor decay with hadrons in $p$+$p$ and 
Au+Au collisions at $\sqrt{s_{NN}}=$200~GeV}

\newcommand{\abilene}{Abilene Christian University, Abilene, Texas 79699, USA}
\newcommand{\acadsin}{Institute of Physics, Academia Sinica, Taipei 11529, Taiwan}
\newcommand{\banaras}{Department of Physics, Banaras Hindu University, Varanasi 221005, India}
\newcommand{\barc}{Bhabha Atomic Research Centre, Bombay 400 085, India}
\newcommand{\bnlcoll}{Collider-Accelerator Department, Brookhaven National Laboratory, Upton, New York 11973-5000, USA}
\newcommand{\bnlphys}{Physics Department, Brookhaven National Laboratory, Upton, New York 11973-5000, USA}
\newcommand{\caucr}{University of California - Riverside, Riverside, California 92521, USA}
\newcommand{\charlesczech}{Charles University, Ovocn\'{y} trh 5, Praha 1, 116 36, Prague, Czech Republic}
\newcommand{\chonbuk}{Chonbuk National University, Jeonju, 561-756, Korea}
\newcommand{\ciae}{China Institute of Atomic Energy (CIAE), Beijing, People's Republic of China}
\newcommand{\cns}{Center for Nuclear Study, Graduate School of Science, University of Tokyo, 7-3-1 Hongo, Bunkyo, Tokyo 113-0033, Japan}
\newcommand{\colorado}{University of Colorado, Boulder, Colorado 80309, USA}
\newcommand{\columbia}{Columbia University, New York, New York 10027 and Nevis Laboratories, Irvington, New York 10533, USA}
\newcommand{\czechtech}{Czech Technical University, Zikova 4, 166 36 Prague 6, Czech Republic}
\newcommand{\dapnia}{Dapnia, CEA Saclay, F-91191, Gif-sur-Yvette, France}
\newcommand{\debrecen}{Debrecen University, H-4010 Debrecen, Egyetem t{\'e}r 1, Hungary}
\newcommand{\elte}{ELTE, E{\"o}tv{\"o}s Lor{\'a}nd University, H - 1117 Budapest, P{\'a}zm{\'a}ny P. s. 1/A, Hungary}
\newcommand{\ewha}{Ewha Womans University, Seoul 120-750, Korea}
\newcommand{\fit}{Florida Institute of Technology, Melbourne, Florida 32901, USA}
\newcommand{\fsu}{Florida State University, Tallahassee, Florida 32306, USA}
\newcommand{\gsu}{Georgia State University, Atlanta, Georgia 30303, USA}
\newcommand{\hiroshima}{Hiroshima University, Kagamiyama, Higashi-Hiroshima 739-8526, Japan}
\newcommand{\ihepprot}{IHEP Protvino, State Research Center of Russian Federation, Institute for High Energy Physics, Protvino, 142281, Russia}
\newcommand{\illuiuc}{University of Illinois at Urbana-Champaign, Urbana, Illinois 61801, USA}
\newcommand{\instpasczech}{Institute of Physics, Academy of Sciences of the Czech Republic, Na Slovance 2, 182 21 Prague 8, Czech Republic}
\newcommand{\isu}{Iowa State University, Ames, Iowa 50011, USA}
\newcommand{\jinrdubna}{Joint Institute for Nuclear Research, 141980 Dubna, Moscow Region, Russia}
\newcommand{\jyvaskyla}{Helsinki Institute of Physics and University of Jyv{\"a}skyl{\"a}, P.O.Box 35, FI-40014 Jyv{\"a}skyl{\"a}, Finland}
\newcommand{\kek}{KEK, High Energy Accelerator Research Organization, Tsukuba, Ibaraki 305-0801, Japan}
\newcommand{\kfki}{KFKI Research Institute for Particle and Nuclear Physics of the Hungarian Academy of Sciences (MTA KFKI RMKI), H-1525 Budapest 114, POBox 49, Budapest, Hungary}
\newcommand{\korea}{Korea University, Seoul, 136-701, Korea}
\newcommand{\kurchatov}{Russian Research Center ``Kurchatov Institute", Moscow, Russia}
\newcommand{\kyoto}{Kyoto University, Kyoto 606-8502, Japan}
\newcommand{\labllr}{Laboratoire Leprince-Ringuet, Ecole Polytechnique, CNRS-IN2P3, Route de Saclay, F-91128, Palaiseau, France}
\newcommand{\lawllnl}{Lawrence Livermore National Laboratory, Livermore, California 94550, USA}
\newcommand{\losalamos}{Los Alamos National Laboratory, Los Alamos, New Mexico 87545, USA}
\newcommand{\lpc}{LPC, Universit{\'e} Blaise Pascal, CNRS-IN2P3, Clermont-Fd, 63177 Aubiere Cedex, France}
\newcommand{\lund}{Department of Physics, Lund University, Box 118, SE-221 00 Lund, Sweden}
\newcommand{\maryland}{University of Maryland, College Park, Maryland 20742, USA}
\newcommand{\mass}{Department of Physics, University of Massachusetts, Amherst, Massachusetts 01003-9337, USA }
\newcommand{\muenster}{Institut fur Kernphysik, University of Muenster, D-48149 Muenster, Germany}
\newcommand{\muhlenberg}{Muhlenberg College, Allentown, Pennsylvania 18104-5586, USA}
\newcommand{\myongji}{Myongji University, Yongin, Kyonggido 449-728, Korea}
\newcommand{\nagasaki}{Nagasaki Institute of Applied Science, Nagasaki-shi, Nagasaki 851-0193, Japan}
\newcommand{\newmex}{University of New Mexico, Albuquerque, New Mexico 87131, USA }
\newcommand{\nmsu}{New Mexico State University, Las Cruces, New Mexico 88003, USA}
\newcommand{\ornl}{Oak Ridge National Laboratory, Oak Ridge, Tennessee 37831, USA}
\newcommand{\orsay}{IPN-Orsay, Universite Paris Sud, CNRS-IN2P3, BP1, F-91406, Orsay, France}
\newcommand{\peking}{Peking University, Beijing, People's Republic of China}
\newcommand{\pnpi}{PNPI, Petersburg Nuclear Physics Institute, Gatchina, Leningrad region, 188300, Russia}
\newcommand{\riken}{RIKEN Nishina Center for Accelerator-Based Science, Wako, Saitama 351-0198, Japan}
\newcommand{\rikjrbrc}{RIKEN BNL Research Center, Brookhaven National Laboratory, Upton, New York 11973-5000, USA}
\newcommand{\rikkyo}{Physics Department, Rikkyo University, 3-34-1 Nishi-Ikebukuro, Toshima, Tokyo 171-8501, Japan}
\newcommand{\saispbstu}{Saint Petersburg State Polytechnic University, St. Petersburg, Russia}
\newcommand{\saopaulo}{Universidade de S{\~a}o Paulo, Instituto de F\'{\i}sica, Caixa Postal 66318, S{\~a}o Paulo CEP05315-970, Brazil}
\newcommand{\seoulnat}{Seoul National University, Seoul, Korea}
\newcommand{\stonybrkc}{Chemistry Department, Stony Brook University, SUNY, Stony Brook, New York 11794-3400, USA}
\newcommand{\stonycrkp}{Department of Physics and Astronomy, Stony Brook University, SUNY, Stony Brook, New York 11794-3400, USA}
\newcommand{\subatech}{SUBATECH (Ecole des Mines de Nantes, CNRS-IN2P3, Universit{\'e} de Nantes) BP 20722 - 44307, Nantes, France}
\newcommand{\tenn}{University of Tennessee, Knoxville, Tennessee 37996, USA}
\newcommand{\titech}{Department of Physics, Tokyo Institute of Technology, Oh-okayama, Meguro, Tokyo 152-8551, Japan}
\newcommand{\tsukuba}{Institute of Physics, University of Tsukuba, Tsukuba, Ibaraki 305, Japan}
\newcommand{\vandy}{Vanderbilt University, Nashville, Tennessee 37235, USA}
\newcommand{\waseda}{Waseda University, Advanced Research Institute for Science and Engineering, 17 Kikui-cho, Shinjuku-ku, Tokyo 162-0044, Japan}
\newcommand{\weizmann}{Weizmann Institute, Rehovot 76100, Israel}
\newcommand{\yonsei}{Yonsei University, IPAP, Seoul 120-749, Korea}
\affiliation{\abilene}
\affiliation{\acadsin}
\affiliation{\banaras}
\affiliation{\barc}
\affiliation{\bnlcoll}
\affiliation{\bnlphys}
\affiliation{\caucr}
\affiliation{\charlesczech}
\affiliation{\chonbuk}
\affiliation{\ciae}
\affiliation{\cns}
\affiliation{\colorado}
\affiliation{\columbia}
\affiliation{\czechtech}
\affiliation{\dapnia}
\affiliation{\debrecen}
\affiliation{\elte}
\affiliation{\ewha}
\affiliation{\fit}
\affiliation{\fsu}
\affiliation{\gsu}
\affiliation{\hiroshima}
\affiliation{\ihepprot}
\affiliation{\illuiuc}
\affiliation{\instpasczech}
\affiliation{\isu}
\affiliation{\jinrdubna}
\affiliation{\jyvaskyla}
\affiliation{\kek}
\affiliation{\kfki}
\affiliation{\korea}
\affiliation{\kurchatov}
\affiliation{\kyoto}
\affiliation{\labllr}
\affiliation{\lawllnl}
\affiliation{\losalamos}
\affiliation{\lpc}
\affiliation{\lund}
\affiliation{\maryland}
\affiliation{\mass}
\affiliation{\muenster}
\affiliation{\muhlenberg}
\affiliation{\myongji}
\affiliation{\nagasaki}
\affiliation{\newmex}
\affiliation{\nmsu}
\affiliation{\ornl}
\affiliation{\orsay}
\affiliation{\peking}
\affiliation{\pnpi}
\affiliation{\riken}
\affiliation{\rikjrbrc}
\affiliation{\rikkyo}
\affiliation{\saispbstu}
\affiliation{\saopaulo}
\affiliation{\seoulnat}
\affiliation{\stonybrkc}
\affiliation{\stonycrkp}
\affiliation{\subatech}
\affiliation{\tenn}
\affiliation{\titech}
\affiliation{\tsukuba}
\affiliation{\vandy}
\affiliation{\waseda}
\affiliation{\weizmann}
\affiliation{\yonsei}
\author{A.~Adare} \affiliation{\colorado}
\author{S.~Afanasiev} \affiliation{\jinrdubna}
\author{C.~Aidala} \affiliation{\mass}
\author{N.N.~Ajitanand} \affiliation{\stonybrkc}
\author{Y.~Akiba} \affiliation{\riken} \affiliation{\rikjrbrc}
\author{H.~Al-Bataineh} \affiliation{\nmsu}
\author{J.~Alexander} \affiliation{\stonybrkc}
\author{K.~Aoki} \affiliation{\kyoto} \affiliation{\riken}
\author{L.~Aphecetche} \affiliation{\subatech}
\author{Y.~Aramaki} \affiliation{\cns}
% ...
\collaboration{PHENIX Collaboration} \noaffiliation

\date{\today}

\begin{abstract}

Measurements of electrons from the decay of open-heavy-flavor mesons 
have shown that the yields are suppressed in Au+Au collisions 
compared to expectations from binary-scaled $p$+$p$ collisions.  These 
measurements indicate that charm and bottom quarks interact with the 
hot-dense matter produced in heavy-ion collisions much more than 
expected.  Here we extend these studies to two-particle correlations 
where one particle is an electron from the decay of a heavy-flavor 
meson and the other is a charged hadron from either the decay of the 
heavy meson or from jet fragmentation.  These measurements provide 
more detailed information about the interactions between heavy quarks 
and the matter, such as whether the modification of the away-side-jet 
shape seen in hadron-hadron correlations is present when the trigger 
particle is from heavy-meson decay and whether the overall level of 
away-side-jet suppression is consistent.  We statistically subtract 
correlations of electrons arising from background sources from the 
inclusive electron-hadron correlations and obtain two-particle 
azimuthal correlations at $\sqrt{s_{NN}}=$200~GeV between electrons 
from heavy-flavor decay with charged hadrons in $p$+$p$ and also 
first results in Au+Au collisions.  We find the away-side-jet shape 
and yield to be modified in Au+Au collisions compared to $p$+$p$ 
collisions.

\end{abstract}

\maketitle

\section{Introduction} \label{sec:intro} 

Experiments at the Relativistic Heavy Ion Collider at Brookhaven 
National Laboratory have produced a hot dense partonic 
matter~\cite{ppg048,ppg086}.  Results from high-\pt $\pi^0$ production 
indicate that fast partons moving through the matter lose a 
substantial amount of energy through interactions~\cite{ppg003,ppg080}.  
This energy loss was expected to be reduced for heavy charm and bottom 
quarks due to the dead cone effect which suppresses gluon 
radiation~\cite{dead_cone_dk}.  However, electrons from the 
semileptonic decay of $D$ and $B$ mesons are seen to be suppressed at 
nearly the same level as $\pi^0$s out to the highest measured $p_T$, 
$\approx$10~GeV/$c$~\cite{ppg066}.  This challenges the picture of 
gluon radiation as the dominant means of parton energy loss.  Various 
alternative scenarios including collisional energy 
loss~\cite{coll_el}, in-medium formation and dissociation of the heavy 
meson~\cite{vitev_dis}, or an increase in the fraction of heavy quarks 
carried by baryons~\cite{sorensen_hq,martinez_hq} have been proposed 
to account for the large suppression.

Single particle yield measurements provide information on the overall 
deviation of particle production from $p$+$p$ expectations.  
However, the observed high-$p_T$ spectra 
are thought to be dominated by particles that have lost less than 
the average amount of energy, either due to a short path 
length through the matter or by a fluctuation.  In order to get more 
detailed information about the interactions between the particles and 
the matter two-particle azimuthal correlations have been extensively 
used.  In $p$+$p$ collisions these correlations are characterized by 
two back-to-back jet peaks~\cite{ppg029}.  At small azimuthal angular 
difference, $\Delta\phi$, particles are from the fragmentation of the 
same jet; at $\Delta\phi\approx\pi$ particles are from the 
fragmentation of partons in the opposing jet.

In heavy-ion collisions, these correlations can provide information 
about the pattern of energy loss for the back-to-back dijet system as 
well as other interactions between the fast partons and the medium.  
Measurements of hadrons associated with a high-\pt hadron have shown 
the away-side correlations from back to back dijets to be 
significantly suppressed~\cite{star_b2b,star_dijets,ppg083,ppg106}.  
Measurements of the correlations of electrons from heavy-flavor decay 
with other hadrons in the event can also provide insight into 
heavy-flavor energy loss and how this compares to $\pi^0$ and direct 
photon triggered correlations where the modifications could be 
different due to the different partons probing the matter.  This is 
crucial for building a quantitative understanding of the nature of 
the interactions between hard partons and the produced hot matter.

In addition, a strong broadening and double peak (shoulder) structure 
of away-side correlations at moderate $p_T$ has been 
observed~\cite{ppg032}.  Many theoretical ideas have been proposed to 
explain this modification, including \v{C}erenkov gluon 
radiation~\cite{cherenkov_dremin, cherenkov_koch}, large angle gluon 
radiation\cite{large_angle_vitev,large_angle_salgado} and Mach 
shock-waves~\cite{casalderrey_mach}.  Measurements of the shoulder 
structure with particles from the fragmentation of heavy quarks, 
especially bottom, are interesting because at moderate momenta the 
quark velocity will be much smaller than the speed of light, in 
contrast to light quarks where $v\approx$c at all jet momenta.  In a 
Mach shock-wave scenario the cone angle of the double peak structure 
away from $\pi$, ($\theta_M$), is related to the speed of the parton 
by: $\cos\theta_M = \frac{c_S}{v}$ where $c_S$ is the speed of sound 
in the matter and $v$ is the speed of the parton as it propagates 
through the matter.  It has been proposed that double-peaked 
correlations that do not obey Mach's Law could favor strongly-coupled 
AdS/CFT string drag scenarios~\cite{betz_dijet}, or transverse 
flow~\cite{betz_flow}.  An alternative explanation based on 
geometrical fluctuations in the initial state leading to triangular 
flow has also recently been proposed~\cite{takahashi,sorensen,alver}.
 
Measurements in $p$+$p$ collisions are a necessary baseline to heavy 
ion measurements, particularly for heavy-flavor triggered 
correlations.  At leading order, several sub-processes contribute to 
charm production leading to a midrapidity $D$ meson.  For 
$p_T<$10~GeV/$c$, one leading order calculation shows 
that $\approx$20\% 
of the time the charm quark leading to the $D$, which decays 
semileptonically into the trigger electron, is balanced by an 
opposing ($\Delta\phi\approx\pi$) $\bar{c}$ quark~\cite{vitev_cnm}. 
The rest of the contribution is from processes such as $cg\to cg$ 
or $cq(\bar{q}) \to cq(\bar{q})$ where the $c$ is not balanced by a 
midrapidity, high-$p_T$ $\bar{c}$.  Next to leading order effects are 
known to be large in heavy quark production.  The {\sc powheg} Monte 
Carlo calcualtion~\cite{powheg}, which includes 2$\to$3 processes, also 
shows substantial contributions to the away-side correlations from gluons.  
Thus, in order to measure $c\bar{c}$ or $b\bar{b}$ correlations one 
should identify the heavy quark in both the trigger and away jets.  
For the present purposes, this means it is not possible to identify 
the jet opposing the electron from heavy-flavor decay unambiguously 
as also from heavy-quark fragmentation.  This complicates the 
interpretation of the present measurements.  However, the comparison 
of heavy- and light-flavor triggered correlations can still provide a 
crucial step toward understanding fast parton propagation through the 
matter.

In the present work heavy-flavor electrons are those from the decay 
of both $D$ and $B$ mesons.  The relative contribution of electrons 
from bottom to the total heavy-flavor electron yield changes with the 
$p_T$ of the electron and has been measured in $p$+$p$ collisions to 
be from $\approx$10--50\% for 
1.0$<p_T<$6.0~GeV/$c$~\cite{ppg094,star_bfrac}.  Fixed-order plus 
next-to-leading-log (FONLL) calculations from Ref.~\cite{fonll} agree well 
with the measured bottom contribution.

We present first results of the azimuthal correlations between 
electrons from the decay of heavy-flavor mesons with charged hadrons 
in Au+Au and $p$+$p$ collisions.  We statistically subtract 
correlations from electrons due to background electron sources 
(Dalitz decays, photon conversions and quarkonia) from the measured 
inclusive electron-hadron correlations.

The paper is organized as follows.  In Section~\ref{analysis_method} 
we outline the analysis procedure used; in Section~\ref{results} we 
show the results in $p$+$p$ and Au+Au collisions; and in 
Section~\ref{conclusions} we conclude and discuss the prospects for 
future measurements.

\section{Analysis Method}
\label{analysis_method}
\subsection{Experimental Setup}

These results are based on 1.1 billion level-1 triggered $p$+$p$ 
events sampling 8.0~$pb^{-1}$ taken during the 2006 RHIC running 
period and 2.6 billion minimum bias Au+Au events, corresponding to 
0.41~$nb^{-1}$ taken during the 2007 RHIC running period.  The events 
were triggered by a hit in each of two Beam-Beam Counters (BBC) at 
3.1$<|\eta|<$3.9 and the interaction is required to be within 25~$cm$ 
of the center of the interaction region.  The $p$+$p$ level-1 
triggered sample also required an energy deposit of approximately 
1.4~GeV in an overlapping tile of 4x4 EMCal towers in coincidence 
with the BBC trigger.  EMCal towers are $\Delta\phi \times 
\Delta\eta\approx 0.01 \times 0.01$.  In Au+Au collisions the event 
centrality is measured by the charge seen in the Beam-Beam Counters 
(BBC)~\cite{phenix_inner}.

The charged particle tracks and photons used in this analysis are 
measured in the PHENIX central arm spectrometers.  Electrons are 
measured between 1.5 and 4.5GeV/$c$ and charged hadrons are measured 
between 0.5 and 4.5GeV/$c$.  PHENIX has two such spectrometers, East 
and West Arms, each covering $\pi/2$~rad in azimuth and $|\eta|<$0.35. 
This analysis uses in each arm a drift chamber (DC), two layers of pad 
chambers (PC1 and PC3), a Ring Imaging \v{C}erenkov Detector (RICH), 
and an electromagnetic calorimeter (EMCal).  Charged particles (both 
electrons and hadrons) are reconstructed in the DC and PC1.  Electron 
identification is done by requiring two (three) associated hits in the 
RICH for $p$+$p$ (Au+Au), a shower shape cut in the EMCal and an $E/p$ 
cut, where $E$ is the energy of the cluster in the EMCal and $p$ is the 
track momentum determined by the DC.  Electron candidates are required 
to have a matching hit in the EMCal within 3$\sigma$ (2$\sigma$) in 
$p$+$p$ (Au+Au).  Cuts on the RICH ring center and shape are also 
included.  The hadron contamination remaining is less than 1\% in 
$p$+$p$ collisions and less than 3\% in Au+Au collisions.  Hadrons are 
identified by a RICH veto and a confirming hit in the PC3.  Photons 
are identified by a shower shape cut in the EMCal and a veto in the 
PC3 to reject charged tracks.  Hadron contamination in the photon 
sample is less than 4\%.  Cuts on electron-hadron and photon-hadron 
pairs are also used to equalize the pair acceptance between real and 
mixed pairs and remove pairs that share hits in the various detector 
subsystems.

In the Au+Au running period the Hadron Blind Detector 
(HBD)~\cite{hbd} was installed for a commissioning run between the 
beam collision vertex position and the central arms.  Photon 
conversions in the detector material were an additional source of 
background electrons in the inclusive electron sample.  The HBD in 
front of the West Arm was absent for a substantial portion of the 
running period.  We make the additional requirement that electrons 
from the 2007 running period are reconstructed in the West Arm and 
select events only from the running period where the HBD in front of 
the West Arm was removed in order to reduce the number of photon 
conversions.

\subsection{Background Subtraction}

In two particle correlations there is a combinatorial background due 
to pairs where the particles are uncorrelated except by event-wise 
correlations, such as centrality and the reaction plane in Au+Au or 
the underlying event in $p$+$p$ collisions.  This background is very 
large in central Au+Au collisions and much smaller in $p$+$p$ 
collisions.  In Au+Au collisions the background is removed by the 
Absolute Background Subtraction technique~\cite{abs} while in $p$+$p$ 
collisions by the Zero Yield at Minimum (ZYAM) method~\cite{zyam} 
(with the uncertainties determined as in Ref.~\cite{abs}) is used to 
subtract the $\Delta\phi$ independent underlying event.  In the 
Absolute Background Subtraction method the combinatorial background 
yield is determined from the centrality dependence of the single 
particle yields.  The principal advantages of this method for this 
analysis are that the background uncertainty is not subject to 
statistical fluctuations caused by the small number of electron-hadron 
pairs and no assumption is made about the shape of the pair 
$\Delta\phi$ distribution.

In $p$+$p$ collisions the underlying event independent of 
$\Delta\phi$.  However, in Au+Au collisions there are additional 
correlations due to elliptic flow, $v_2$.  These correlations do not 
affect the magnitude of the combinatorial background, but do affect 
the azimuthal correlation of particles.  The $v_2$ values used in this 
analysis are measured using the reaction plane as determined from the 
PHENIX Reaction Plane Detector and the same particle cuts used in the 
correlation analysis.

Results are reported as conditional yields of hadrons associated with 
trigger electrons after combinatorial background subtraction, 
$Y_{e-h}(\Delta\phi)$:
\begin{equation}
Y_{e-h}(\Delta\phi) = 
\frac{1}{N_{e}\epsilon_h}
\frac{dN_{\rm meas}}{d\Delta\phi} - B(1+2v_2^{e}v_2^{h}\cos(2\Delta\phi)) 
\end{equation}
where $N_{e}$ is the total number of observed trigger particles and 
$\epsilon_h$ is the reconstruction efficiency for the associated 
hadrons as determined by a {\sc geant} based Monte Carlo simulation 
and embedding single particles into real events.  $\frac{dN_{\rm 
meas}}{d\Delta\phi}$ is the measured trigger-associated particle 
$\Delta\phi$ distribution, which has been corrected for nonuniform two 
particle $\Delta\phi$ acceptance by using mixed events~\cite{abs}.  
$B$ is determined by the background subtraction methods described 
above.

\subsection{Removal of Non-Open Heavy-Flavor Electron-Hadron Correlations}

Studies of electrons from open heavy-flavor decay are complicated by 
the background of electrons from light meson decay, photon 
conversions and, at higher $p_T$, quarkonia and Drell-Yan.  In this 
analysis we statistically subtract the correlations from these 
background sources using a method similar to that used to measure 
direct photon-hadron correlations~\cite{ppg090}.  The yield of 
inclusive electron-hadron pairs per electron trigger, the conditional 
yield $Y_{e_{inc}-h}(p_{T,e},p_{T,h},\Delta\phi)$, is measured.  
This is a weighted average of the conditional yield of hadrons 
associated with electrons from heavy-flavor decay and the conditional 
yield of hadrons associated with electrons from the background 
sources (the dependencies on $p_T$ and $\Delta\phi$ are omitted 
hereafter for simplicity):
\begin{equation}
Y_{e_{inc}-h}
= \frac{N_{e_{\rm HF}} Y_{e_{\rm HF}-h}
+ N_{e_{bkg}} Y_{e_{bkg}-h}}
{N_{e_{\rm HF}} + N_{e_{bkg}}}
\end{equation}
where $N_{e_{\rm HF}}$ ($N_{e_{bkg}}$) is the number of electrons 
from heavy-flavor decay (background sources).  $Y_{e_{\rm HF}-h}$ can 
then be written as:
\begin{equation}
Y_{e_{\rm HF}-h} = \frac{(R_{\rm HF}+1)Y_{e_{inc}-h} - Y_{e_{bkg}-h}}{R_{\rm HF}}
\label{sub_eq}
\end{equation}
where $R_{\rm HF}\equiv \frac{N_{e_{\rm HF}}}{N_{e_{bkg}}}$.  The two 
quantities to be determined are $R_{\rm HF}$ and $Y_{e_{bkg}-h}$.  
$R_{\rm HF}$ can be determined by comparing the measured electron 
yields to the known sources of background electrons.  The 
$Y_{e_{bkg}-h}$ determination is based on measured and simulated 
azimuthal correlations of the sources of background electrons and is 
described in detail below.

The $R_{\rm HF}$ value giving the composition of the electron 
triggers into heavy-flavor and background sources is taken from the 
published PHENIX measurements~\cite{ppg065,ppg066} for the electron 
$p_T$ bins used in this analysis.  Based on simulations the 
$R_{\rm HF}$ has been decreased to account for extra air conversions due 
to the removal, in both the $p$+$p$ and Au+Au data samples, of the 
helium bag, which was installed during the data taking periods of 
Refs.~\cite{ppg066,ppg065}.  The removal of the He bag added 0.65\% 
of a radiation length to the material in front of the tracking 
system.  However, the reconstruction efficiency of the electrons from 
these air conversions decreases with the distance from the 
interaction point.  This was simulated using a {\sc geant}-based 
description of the PHENIX detector and the electron-identification 
cuts described above.  Additionally, electrons due to quarkonia 
decay are included in the background sample~\cite{ppg077}, further 
reducing $R_{\rm HF}$.  The reduction due to quarkonia is determined 
by $J/\Psi$ measurements and simulations of the 
decay~\cite{ppg069,ppg068,ppg077}.  The change to the background yield 
due to quarkonia is 3\% at 1.5$<p_T<$2.0~GeV/$c$ and 38\% at 
4.0$<p_T<$4.5~GeV/$c$ in $p$+$p$ collisions (at moderate and 
high $p_T$ the heavy-flavor signal is larger than the background, so the 
change to the heavy-flavor electron spectra is much 
smaller~\cite{ppg077}).  At all but the highest $p_T$ used here the 
change is only due to electrons from $J/\Psi$.  Electrons from 
$\Upsilon$ become important at $p_T>$4.0~GeV/$c$ and electrons from 
Drell-Yan are negligible at all transverse momenta used in this 
analysis.  The $R_{\rm HF}$ values used in this analysis are shown in 
Table~\ref{rhf_pp} ($p$+$p$) and Table~\ref{rhf_AuAu} (Au+Au).

\begin{table}
\caption{Ratio of heavy-flavor electrons to  
background electrons in $p$+$p$ along with the systematic uncertainty.}
\begin{ruledtabular}  \begin{tabular}{cccc}
& $p_T$ (GeV/$c$) & $R_{\rm HF}$ & \\
\hline
& 1.5-2.0 & 0.66 $\pm$ 0.13 & \\
& 2.0-2.5 & 0.86 $\pm$ 0.14 & \\
& 2.5-3.0 & 1.09 $\pm$ 0.14 & \\
& 3.0-3.5 & 1.31 $\pm$ 0.16 & \\
& 3.5-4.0 & 1.49 $\pm$ 0.17 & \\
& 4.0-4.5 & 1.60 $\pm$ 0.17 & \\
\end{tabular} \end{ruledtabular}
\label{rhf_pp}
\end{table}

\begin{table}
\caption{Ratio of heavy-flavor electrons to background 
 electrons in Au+Au along with the systematic uncertainty.}
\begin{ruledtabular} \begin{tabular}{cccc}
& $p_T$ (GeV/$c$) & $R_{\rm HF}$ &  \\
\hline
& 1.5-2.0 & 0.94 $\pm$ 0.21 & \\
& 2.0-2.5 & 1.14 $\pm$ 0.25 & \\
& 2.5-3.0 & 1.29 $\pm$ 0.30 & \\
& 3.0-3.5 & 1.38 $\pm$ 0.34 & \\
& 3.5-4.0 & 1.43 $\pm$ 0.36 & \\
& 4.0-4.5 & 1.38 $\pm$ 0.36 & \\
\end{tabular} \end{ruledtabular}
\label{rhf_AuAu}
\end{table}

The remaining unknown in Eq.~\ref{sub_eq} is the azimuthal 
correlations of the background electrons with hadrons, 
$Y_{e_{bkg}-h}$.  These pairs can be divided into two classes: those 
from {\it photonic} sources, electrons from the decay of light mesons 
and photon conversions in the detector material, and those from 
quarkonia (electrons from light vector meson decay are a small 
contribution and are neglected).  These correlations are determined 
from inclusive photon-hadron correlations.  Inclusive photons, like 
photonic electrons, are largely from $\pi^0$ and $\eta$ decay at 
$p_T<$5~GeV/$c$, and {\sc pythia} (version 6.421)~\cite{pythia} 
simulations of the correlations between electrons from $J/\Psi$ decay 
with hadrons.  The fraction of hadrons misidentified as electrons is 
small, as discussed above, and those correlations are not separately 
subtracted.  At $p_{T,e}<$2~GeV/$c$ there is a small ($\approx$3\%) 
contribution from electrons from the semileptonic decay of kaons 
whose correlations are also neglected.

\paragraph{Photonic Electron Correlations} 

Photonic electron sources include Dalitz decays and photon 
conversions where the photons are from light meson decay.  To 
determine these correlations, $Y_{e_{\rm phot}-h}$, we measure 
inclusive photon-hadron correlations.  Inclusive photons are 
dominantly also from light meson decay.  However, the parent meson 
$p_T$ distributions need not be the same for photonic electrons and 
inclusive photons.  The relationship can be written as:
\begin{equation}
Y_{e_{\rm phot}-h}(p_{T,i}) = \sum_j\ w_i(p_{T,j})\ Y_{\gamma_{inc}-h}(p_{T,j})
\end{equation}
where each $i$ ($j$) represents a 0.5~GeV/$c$ bin in electron 
(photon) $p_T$.  The weight coefficients $w_i(p_{T,j})$ are 
determined via simulation and are used to transform the inclusive 
photon-hadron correlations into expectations for photonic 
electron-hadron correlations.

Two methods are used to determine the $w_i(p_{T,j})$.  
The first method treats the electrons as coming from photon 
conversions in the detector material and the second treats the 
electrons as coming from Dalitz decays.  True photonic electrons come 
from both sources, however both methods give very similar 
$w_i(p_{T,j})$ values.

In the first method the measured single inclusive photon spectrum is 
input into a {\sc geant} based simulation of the PHENIX detector.  
The same electron identification cuts as in the real data analysis 
are then applied to reconstructed conversion electrons and the 
relationship between the input photon $p_T$ and the reconstructed 
conversion electron $p_T$ determines $w$.

In the second method, the $\pi^0$ spectrum from Ref.~\cite{ppg024} 
(for $p$+$p$ collisions) or Ref.~\cite{ppg080} (for Au+Au collisions) 
is taken as input to a Monte Carlo simulation that decays the 
$\pi^0$s via Dalitz decay.  The relationship between the intermediate 
low mass virtual photon and the resulting decay electron are used in 
the same manner as in the first method to determine $w$.  Since the 
mass of the virtual photon is small the difference in the $p_T$ 
distribution between real photons and the virtual photons in the 
Dalitz decay is negligible.

\begin{figure}[bh]
\centering
\includegraphics[width=1.0\linewidth]{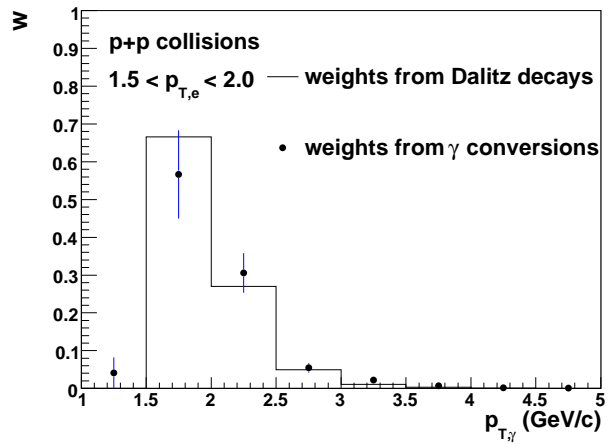}
\caption{(color online) 
Weights ($w$) for electrons with 1.5$<p_T<$2.0~GeV/$c$ from the 
method using Dalitz decays (histogram) and photon conversions 
(solid points) as a function of the photon $p_T$.}
\label{weights}
\end{figure}

Figure~\ref{weights} compares the $w$ from the two methods for a single 
electron $p_T$ selection and shows that the differences between the two 
methods are small.  The maximum deviation in the resulting $e_{\rm HF}$-h 
conditional yields is at small $\Delta\phi$, where the difference 
between the two methods is 0.008 (0.006) in $p$+$p$ (Au+Au).  Both 
inclusive photons and electrons from Dalitz decays are 
largely the result of $\pi^0$ decay.  The $\pi^0$ spectrum falls 
steeply with $p_T$ and thus the measured photonic electrons at a 
given $p_T$ are dominated by those carrying a large fraction of the 
$\pi^0$ $p_T$ regardless of whether the electron comes from a 
conversion or a Dalitz decay.  This argument holds for all heavier 
mesons that contribute to the photonic electron sample, which 
explains the small difference between the conversion method 
(including all mesons that decay into photons) and the Dalitz 
decay method (including only $\pi^0$ decay).

\begin{figure}
\centering
\includegraphics[width=1.0\linewidth]{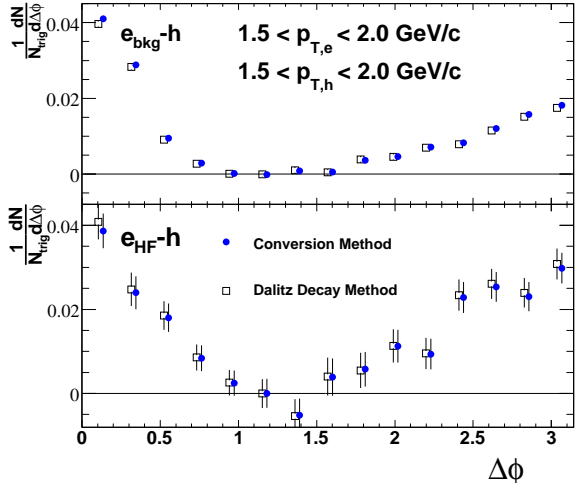}
\caption{(color online) 
$e_{bkg}-h$ (top panel) and $e_{\rm HF}-h$ (bottom panel) 
conditional yields for $p$+$p$ collisions with 
1.5$<p_{T,e}<$2.0~GeV/$c$ and 1.5$<p_{T,h}<$2.0~GeV/$c$ for the two 
methods of constructing $e_{bkg}-h$ conditional yields: the 
conversion method (solid circles) and the Dalitz decay method (black 
squares).  Points have been offset slightly for clarity.}
\label{dalitz_comp}
\end{figure}

Figure~\ref{dalitz_comp} shows the difference in the $e_{bkg}-h$ and 
$e_{\rm HF}-h$ correlations for the two methods.  The difference is 
small compared to the statistical uncertainty and is included in the 
systematic uncertainty.  An additional small systematic uncertainty 
is included from the statistical uncertainty of the simulations.

\begin{figure}[ht]
\includegraphics[width=1.0\linewidth]{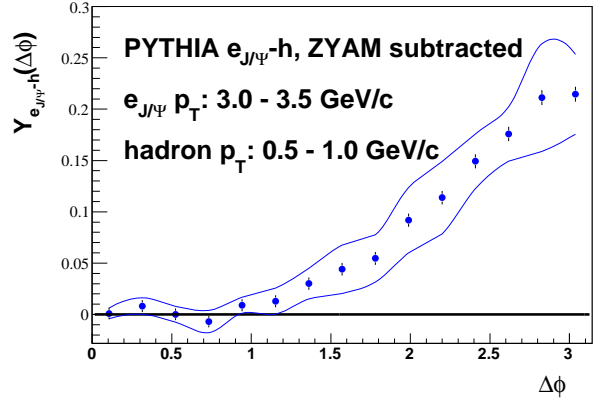}
\caption{(color online) 
{\sc pythia} $e_{J/\Psi}-h$ correlations after ZYAM 
subtraction for electrons with 3.0$<p_T<$3.5~GeV/$c$ and hadrons with 
0.5$<p_T<$1.0~GeV/$c$.  The central values are from the default {\sc 
pythia} $J/\Psi$ production setting and the lines show the systematic 
uncertainty set by the magnitude of the maximal deviation between the 
default setting at the color singlet and color octet production 
settings.}
\label{jpsi_corr}
\end{figure}

\paragraph{Correlations of Electrons from Quarkonia Decay} 

The azimuthal correlations between $J/\Psi$s and hadrons have not yet 
been measured at these momenta, so {\sc pythia}~\cite{pythia} is used 
to simulate the correlations between the electrons from $J/\Psi$ 
decay with charged hadrons.  For both the Au+Au and $p$+$p$ 
measurements the default $J/\Psi$ production within {\sc pythia} is 
used.  For $p$+$p$, the systematic uncertainty is taken as the 
maximal deviation from the default production when varying the 
production mechanism between color singlet ({\sc pythia} ISUB=421) 
and color octet ({\sc pythia} ISUB=422) states.  
Figure~\ref{jpsi_corr} shows the correlations of electrons from 
$J/\Psi$ decay and hadrons after ZYAM background subtraction for an 
example $p_{T}$ selection.  For Au+Au, the situation is more 
uncertain as a substantial fraction of the $J/\Psi$s could be coming 
from recombining $c$ and $\bar{c}$ quarks~\cite{pbm_stat,thews_reco}.  
In this case, the azimuthal correlations could potentially be 
strongly reduced; the systematic uncertainty is taken to extend from 
the {\sc pythia} expectation to no correlation between the decay 
electron and other hadrons.

\section{Results \& Discussion}
\label{results}
\begin{figure*}[ht]
\begin{minipage}{1.0\linewidth}
\includegraphics[width=0.48\linewidth]{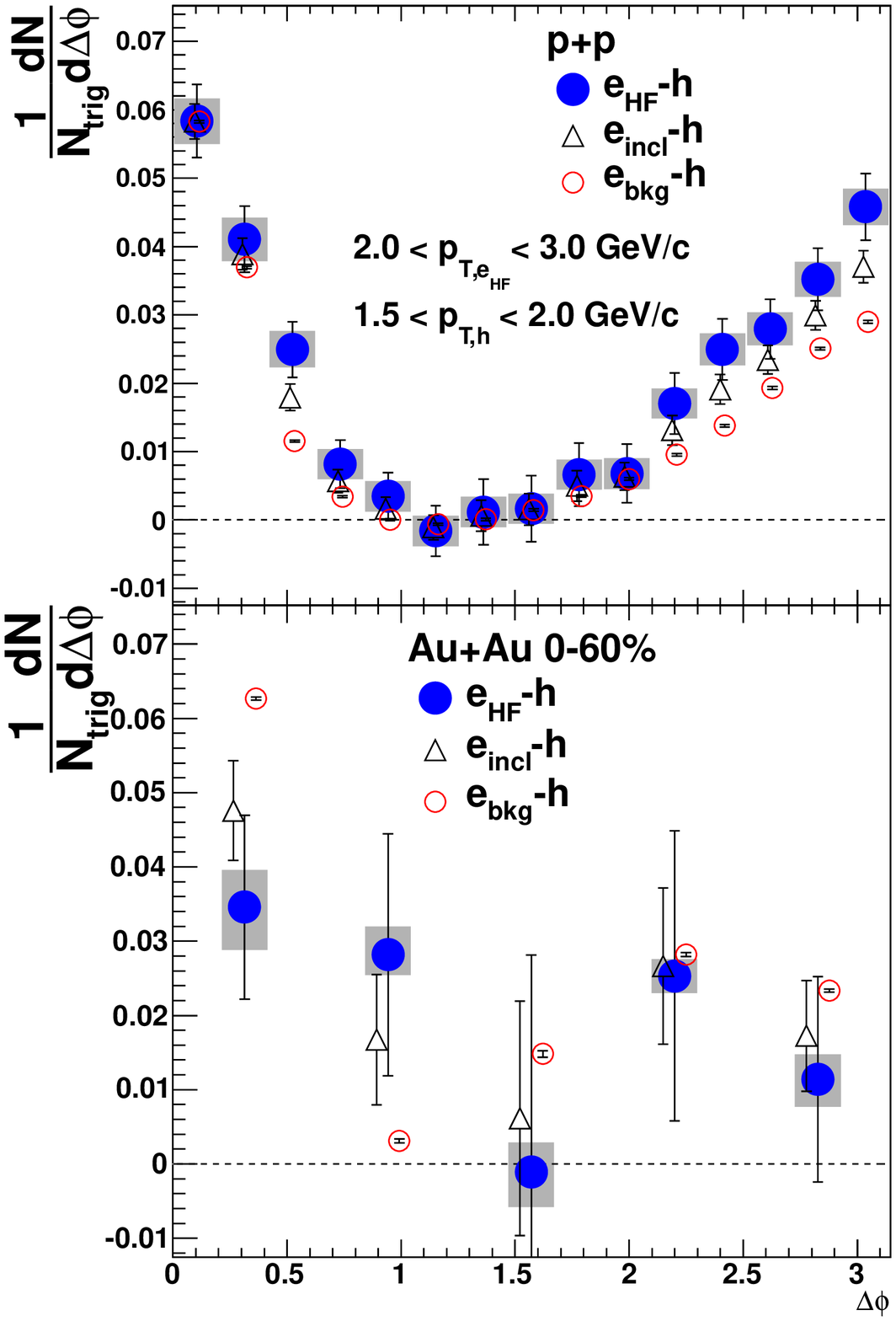}
\caption{(color online) 
$e_{inc}-h$, $e_{bkg}-h$ and $e_{\rm HF}-h$ (solid circles) 
for $p$+$p$ (top panel) and Au+Au (bottom panel) collisions for 
2.0$<p_{T,e}<$3.0~GeV/$c$ and 1.5$<p_{T,h}<$2.0~GeV/$c$.  The overall 
normalization uncertain of 7.9\% in $p$+$p$ and 9.4\% in Au+Au is not 
shown.}
\label{corrfunc}
\end{minipage}

\begin{minipage}{1.0\linewidth}
\includegraphics[width=0.48\linewidth]{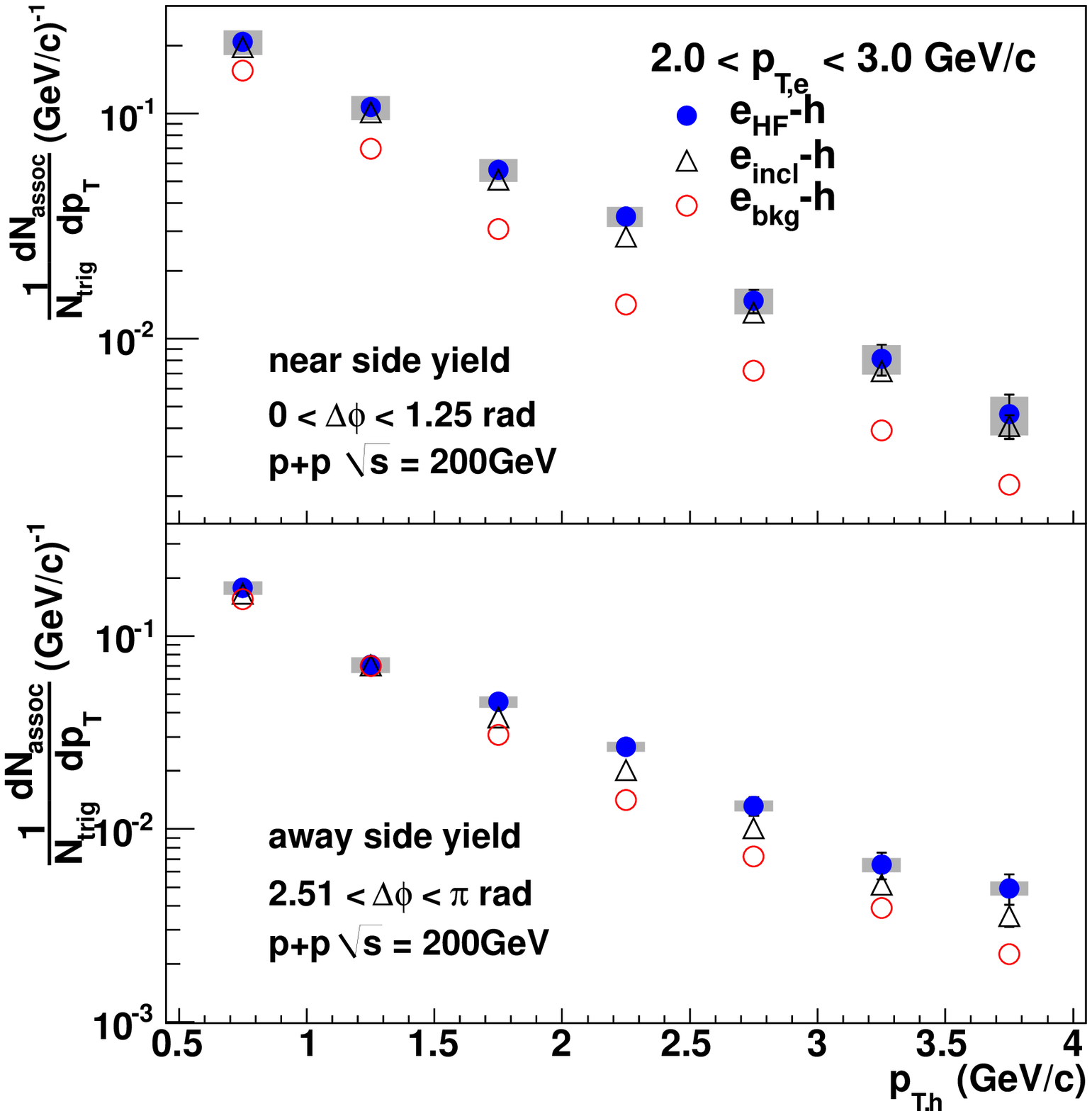}
\caption{(color online) 
Near (top) and away (bottom) side conditional yields in 
$p$+$p$ collisions as a function of hadron $p_T$.  $e_{inc}$ triggers 
are shown as triangles, $e_{bkg}$ triggers are shown as open circles 
and $e_{\rm HF}$ triggers are shown as solid circles.  The boxes on 
the $e_{\rm HF}-h$ points are the systematic uncertainties except for 
the overall normalization uncertainty of 7.9\%, which is not shown.}
\label{eh_pp_yield}
\end{minipage}
\end{figure*}

Example jet functions, after efficiency corrections and combinatorial 
background subtraction, are shown in Fig.~\ref{corrfunc} for 
$e_{inc}$, $e_{bkg}$ and $e_{\rm HF}$ triggers.  Both the near- and 
away-side jet shapes are clearly present in the $p$+$p$ data, however 
the statistical uncertainties in the Au+Au data are much larger.  
The boxes show the systematic uncertainties from all sources except 
for the overall normalization uncertainty of 7.9\% in $p$+$p$ and 
9.4\% in Au+Au.  These jet functions, and others shown in 
Appendix~\ref{jf_appendix}, are integrated and fit to extract the 
yields and widths that follow.  Examples of the integrated near- and 
away-side conditional yields in $p$+$p$ collisions are shown in 
Fig.~\ref{eh_pp_yield} for the $e_{inc}$, $e_{bkg}$ and 
$e_{\rm HF}$ triggers as a function of the hadron $p_T$.

\begin{figure}[hb]
\centering
\includegraphics[width=1.0\linewidth]{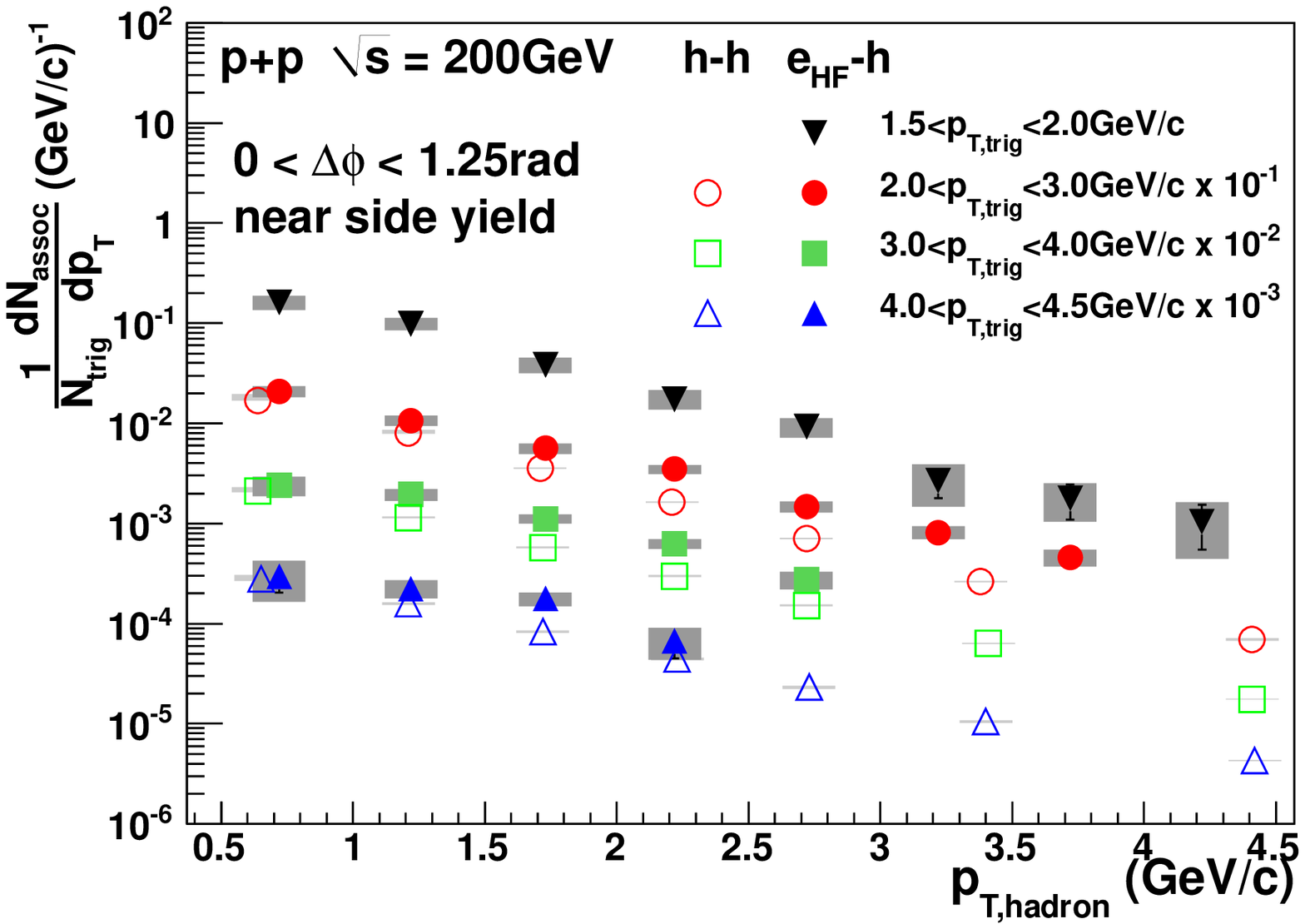}
\caption{(color online) 
Near-side $e_{\rm HF}-h$ conditional yields in $p$+$p$ 
collisions (solid points) as a function of the associated hadron 
$p_T$.  For comparison hadron-hadron conditional yields in $p$+$p$ 
collisions from Ref.~\cite{ppg083} are also shown (the $\Delta\phi$ 
range for the hadron-hadron yields is $\Delta\phi<\pi/3$).  The 
associated hadron $p_T$ spectra are harder for $e_{\rm HF}-h$ than 
hadron-hadron conditional yields at the same $p_{T,trig}$ range (the 
highest $p_T$ trigger selection is for 4.0--4.5~GeV/$c$ for the 
$e_{\rm HF}$ triggers and 4.0--5.0~GeV/$c$ for the hadron triggers).  
The overall normalization uncertainty of 7.9\% is not shown.}
\label{pp_near_yield}
\end{figure}

\subsection{$p$+$p$ Collisions}

\begin{figure}[hb]
\centering
\includegraphics[width=0.97\linewidth]{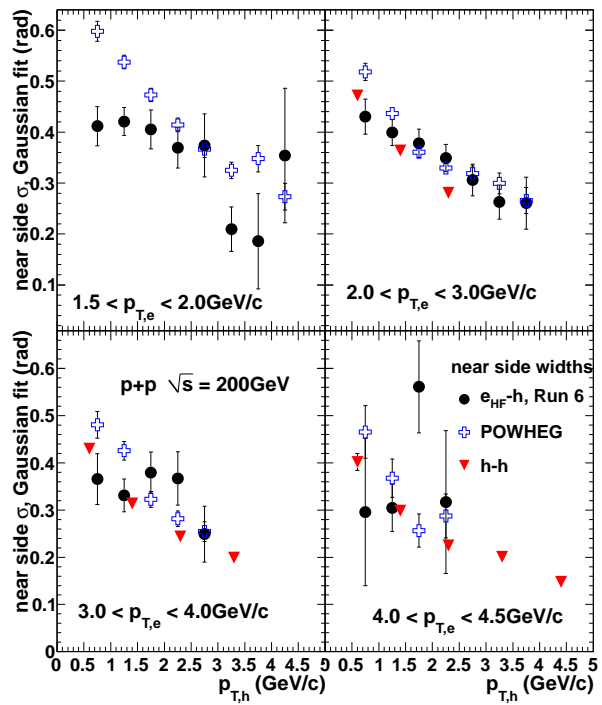}
\caption{(color online) 
Gaussian widths of near-side conditional yields as a 
function of the associated hadron $p_T$ for four $p_{T,e}$ 
selections.  Solid circles show results from $e_{\rm HF}-h$ 
correlations and triangles show results from $h-h$ correlations.  
Crosses are from {\sc powheg}~\cite{powheg} with charm and bottom 
combined according to FONLL calculations~\cite{fonll}.}
\label{widths}
\end{figure}

The near-side (0$<\Delta\phi<$1.25 rad) conditional yields of hadrons 
associated with heavy-flavor electrons are shown in 
Fig.~\ref{pp_near_yield} for the four electron $p_T$ selections used 
in this analysis.  While, in general, parton fragmentation favors the 
production of hadrons carrying a small fraction of the parent quark 
momentum, the heavy meson, $D$ or $B$, resulting from a heavy quark 
fragmentation typically carries a large fraction of the heavy quark 
momentum ($z=\frac{p_{hadron}}{p_{\rm jet}}$ is peaked at 
$\approx$0.60 for charm and $\approx$0.85 for 
bottom)~\cite{cleoc,bellec,alephb,opalb,sldb}.  For comparison we also 
show the conditional yields from correlations between two charged 
hadrons from Ref.~\cite{ppg083}.  The spectra of near-side associated 
hadrons is harder for the $e_{\rm HF}$ triggers than for the hadron 
triggers in all overlapping $p_{T,trig}$ selections.  The near-side 
correlations are expected to be dominated by hadrons that are also 
from the decay of the heavy meson.  The large mass of the heavy meson 
translates to a wider expected near-side correlation when the hadron 
and the electron are both from the heavy meson decay.  
Figure~\ref{widths} shows the Gaussian widths of the near-side 
conditional yields as a function of the associated hadron $p_T$.  
Also shown for comparison are the near-side widths for hadron-hadron 
correlations~\cite{ppg083}, which primarily come from light parton 
jets.  The widths of the $e_{\rm HF}-h$ correlations are slightly 
wider for 2.0$<e_{\rm HF}<$3.0~GeV/$c$, consistent with the near side 
being dominated by decay induced correlations.  For higher $p_T$ 
electrons the statistical uncertainties become too large to make a 
quantitative statement.  Results from {\sc powheg}~\cite{powheg}, a 
next-to-leading-order Monte Carlo calculation, with charm and bottom 
contributions set by FONLL calculations~\cite{fonll} are shown.  
These simulations are consistent with the data except for the lowest 
electron and hadron momenta.

\begin{table*}[ht]
\caption{Mean transverse momentum of the parent $D$ and $B$ mesons 
contributing to the heavy-flavor electron $p_T$ bins used here.  
They are combined according to the fraction of heavy-flavor electrons 
from $b$ quarks, $\frac{b \to e} {(c\to e + b\to e)}$ according to 
the FONLL calculations~\cite{fonll} (as shown in Ref.~\cite{ppg094}) 
to determine the mean heavy meson transverse momentum.}
\begin{ruledtabular} \begin{tabular}{ccccc}
$ p_{T,e}$ (GeV/$c$) & $\langle p_T \rangle_{D}$ (GeV/$c$)&
$\langle p_T \rangle_{B}$ (GeV/$c$) & $\frac{b\to e}{(c\to e + b\to e)}$ & $\langle p_T \rangle_{meson}$ (GeV/$c$)\\
\hline
1.5-2.0 & 3.4 & 4.4 & 0.15 & 3.6\\
2.0-3.0 & 4.1 & 4.7 & 0.26 & 4.3\\
3.0-4.0 & 5.6 & 5.6 & 0.42 & 5.6\\
\end{tabular} \end{ruledtabular}
\label{tabpt}
\end{table*}

\begin{figure}[ht]
\centering
\includegraphics[width=1.0\linewidth]{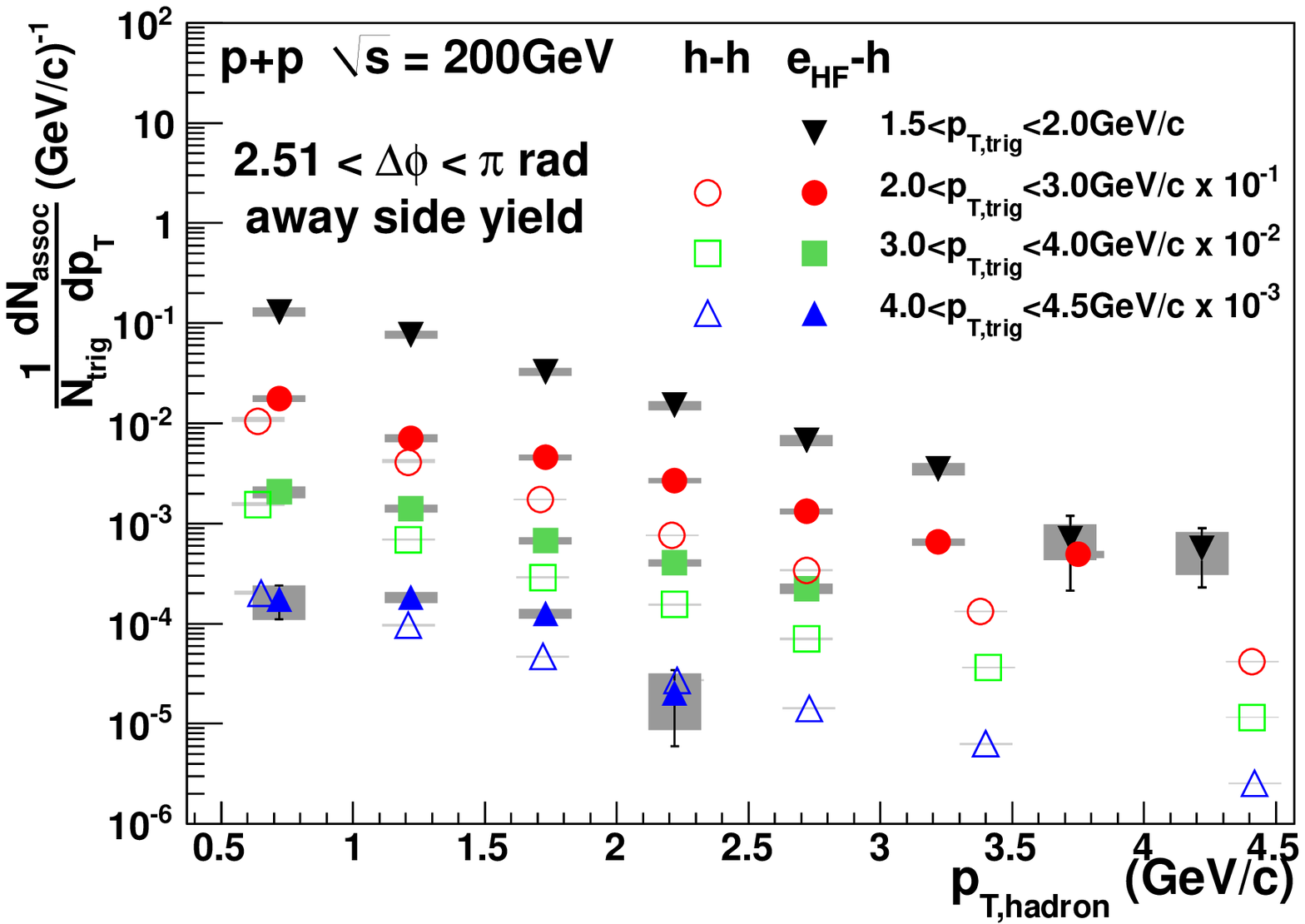}
\caption{(color online) 
Away-side $e_{\rm HF}-h$ (2.51$<\Delta\phi<\pi$) conditional 
yields in $p$+$p$ collisions (solid points) as a function of the 
associated hadron $p_T$.  For comparison hadron-hadron conditional 
yields in $p$+$p$ collisions from Ref.~\cite{ppg083} are also shown 
(the $\Delta\phi$ range for the hadron-hadron yields is 
2.51$\Delta\phi<\pi$).  The associated hadron $p_T$ spectra are 
harder for $e_{\rm HF}-h$ than hadron-hadron conditional yields at 
the same $p_{T,trig}$ range (the highest $p_T$ trigger selection is 
for 4.0--4.5~GeV/$c$ for the $e_{\rm HF}$ triggers and 4.0--5.0~GeV/$c$ 
for the hadron triggers).  The overall normalization uncertainty of 
7.9\% is not shown.}
\label{pp_away_yield}
\end{figure}

\begin{figure}[ht]
\centering
\includegraphics[width=1.0\linewidth]{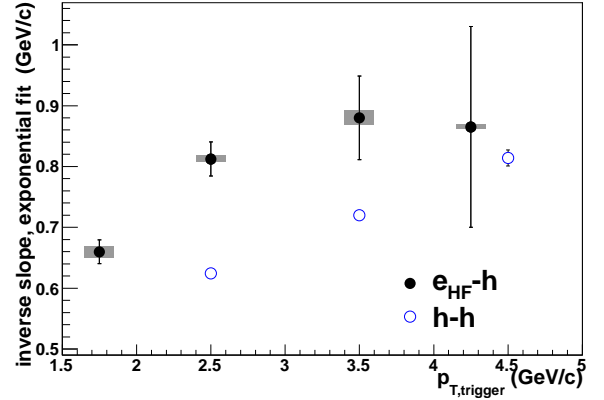}
\caption{(color online) 
Inverse slope of the $e_{\rm HF}$ triggered away-side 
conditional hadron $p_T$ distributions shown in 
Fig.~\ref{pp_away_yield} as a function of $p_T$ of the trigger 
particle.  For comparison fits to the hadron-hadron data from 
Ref.~\cite{ppg083} are also shown.}
\label{slope}
\end{figure}

The away-side (summed over 2.51$<\Delta\phi<\pi$ rad) conditional 
yields are shown in Fig.~\ref{pp_away_yield}.  The yields on the 
away side are dominated by the fragmentation and decay of particles 
in the opposing jet.  As discussed above the opposing dijet does not 
have to contain a balancing heavy-flavor quark; here the yields are a 
mix of heavy and light parton jets.  At a given $p_T$ trigger bin the 
heavy-flavor electron triggered away-side spectrum is harder than the 
light hadron triggered one.  In order to quantify the slope 
differences between light hadron triggers and heavy-flavor electron 
triggers we plot the inverse slope from an exponential fit of the 
away-side spectra in Fig.~\ref{slope}.

However, the electron only carries a fraction of the heavy meson 
$p_T$.  {\sc pythia} was used to estimate the parent meson average 
$p_T$ for both charm and bottom mesons and the results are shown in 
Table~\ref{tabpt}.  When comparing the inverse slopes at similar 
meson $p_T$, as opposed to similar trigger particle $p_T$, the 
difference in the inverse slopes between the two trigger types 
largely disappears.

\begin{figure*}[ht]
\begin{minipage}{0.48\linewidth}
\includegraphics[width=0.99\linewidth]{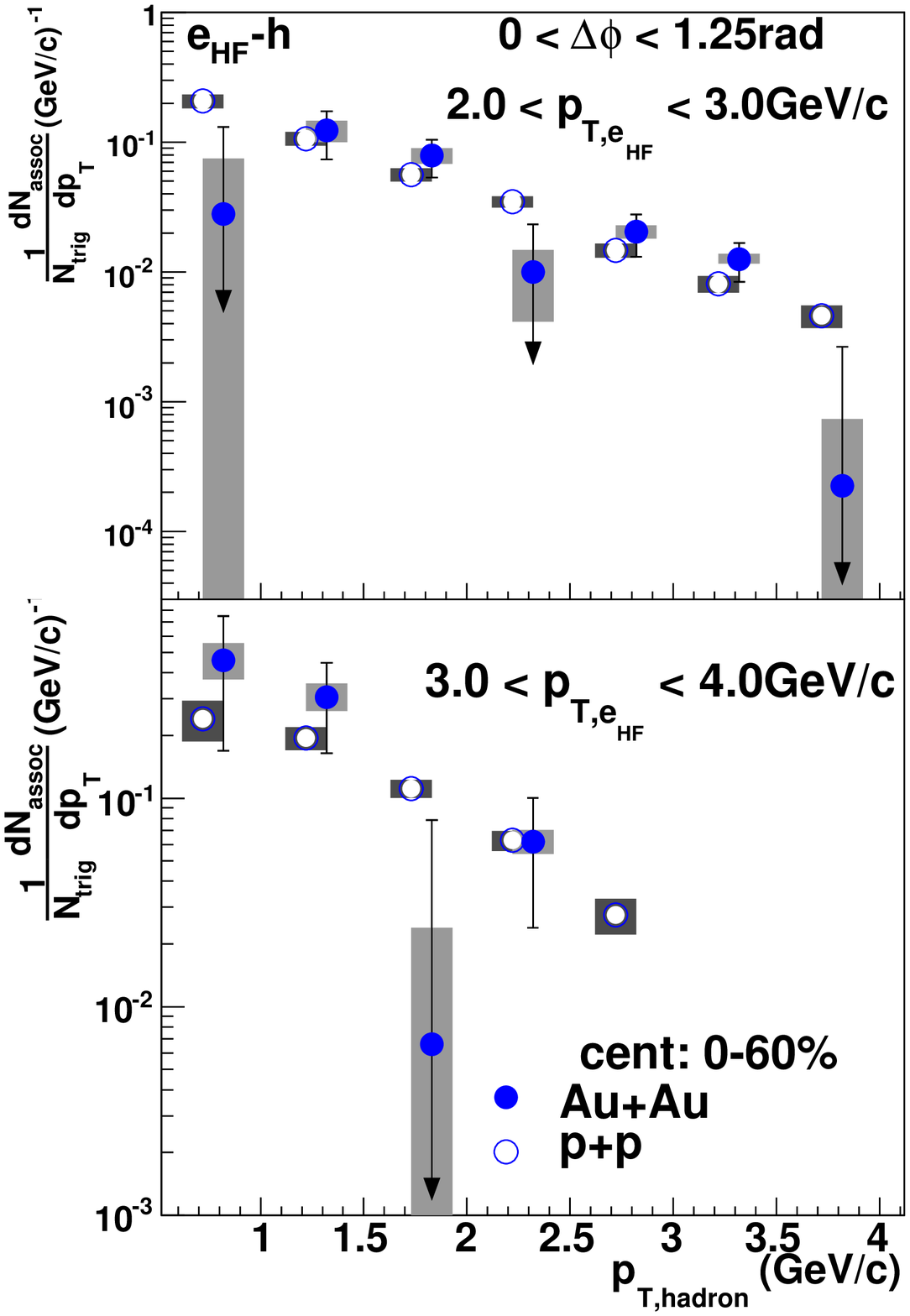}
\caption{\label{nearyield_auau} (color online) 
Near-side (0$<\Delta\phi<$1.25 rad) integrated yield for 
Au+Au (solid circles) and $p$+$p$ collisions (open circles) for 
2.0$<p_{T,e}<$3.0~GeV/$c$ (top panel) and 3.0$<p_{T,e}<$4.0~GeV/$c$ 
(bottom panel) as a function of the associated hadron $p_T$.  The 
overall normalization uncertainty of 9.4\% in Au+Au and 7.9\% in 
$p$+$p$ is not shown.  Points are slightly shifted horizontally for 
clarity.}
\end{minipage}%
\hspace{0.5cm}
\begin{minipage}{0.48\linewidth}
\includegraphics[width=0.99\linewidth]{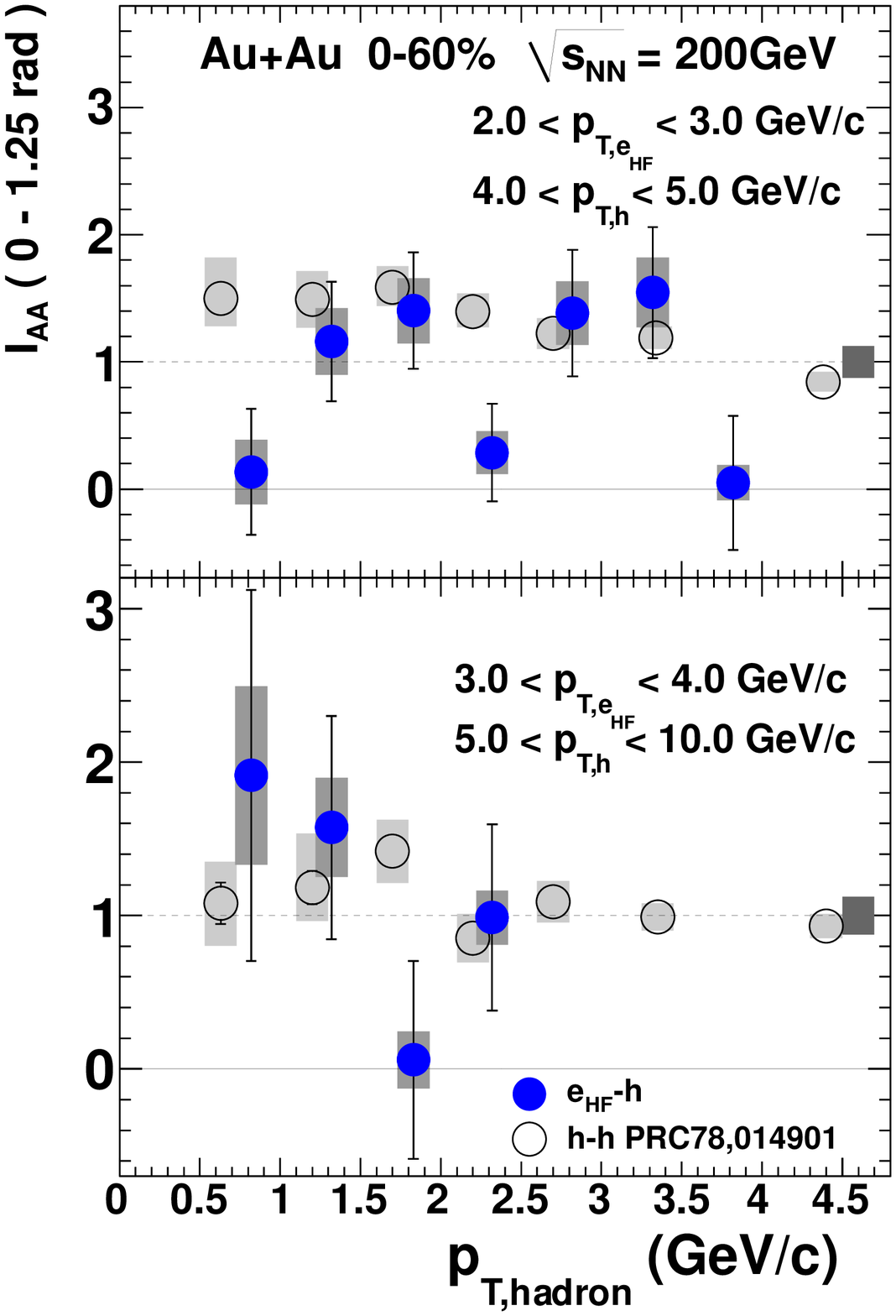}
\caption{\label{iaa_near} (color online) 
Near-side (0$<\Delta\phi<$1.25 rad) $I_{\rm AA}$ for 
2.0$<p_{T,e}<$3.0~GeV/$c$ (top panel) and 3.0$<p_{T,e}<$4.0~GeV/$c$ 
(bottom panel) as a function of the associated hadron $p_T$ for 
$e_{\rm HF}$ (solid points) and hadron (open points) triggers (from 
Ref.~\cite{ppg083}).  The gray band around unity shows the overall 
normalization uncertainty (12.4\%), which moves all points together.  
Points are slightly shifted horizontally for clarity.}
\end{minipage}%
 \end{figure*}

\begin{figure*}[ht]
\begin{minipage}{1.0\linewidth}
\includegraphics[width=0.55\linewidth]{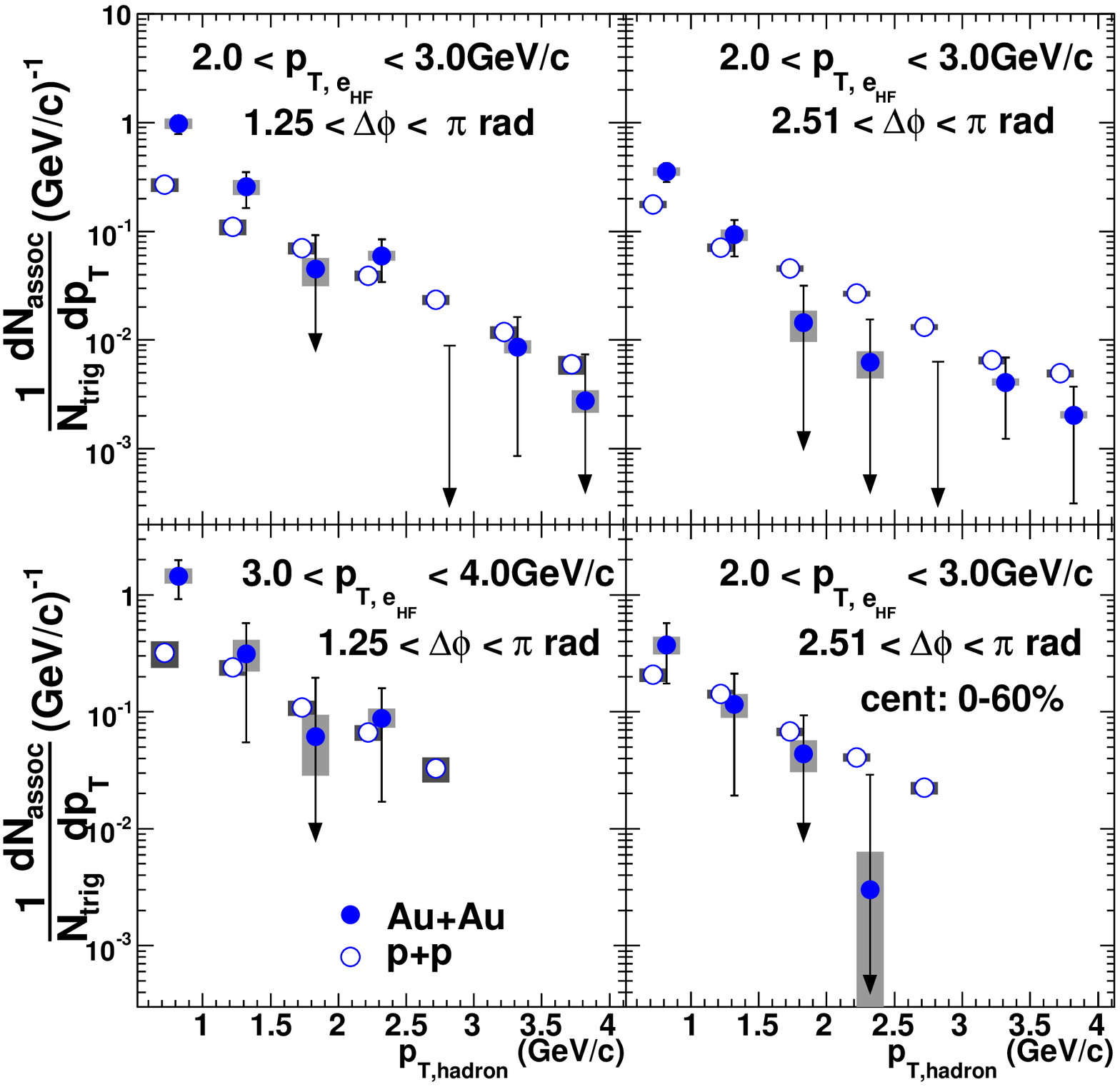}
\caption{\label{yields_away_2panel} (color online) 
Away-side conditional yields for wide (left) and narrow 
(right) away-side $\Delta\phi$ integration ranges for Au+Au (solid 
points) and $p$+$p$ (open points).  Top panels show 
2.0$<p_{T,e}<$3.0~GeV/$c$ and bottom panels shown 
3.0$<p_{T,e}<$4.0~GeV/$c$.  Upper limits are for 90\% confidence 
levels.  The overall normalization uncertainty of 9.4\% in Au+Au and 
7.9\% in $p$+$p$ are not shown.  Points are slightly shifted 
horizontally for clarity.}
\end{minipage}
\begin{minipage}{1.0\linewidth}
\includegraphics[width=0.7\linewidth]{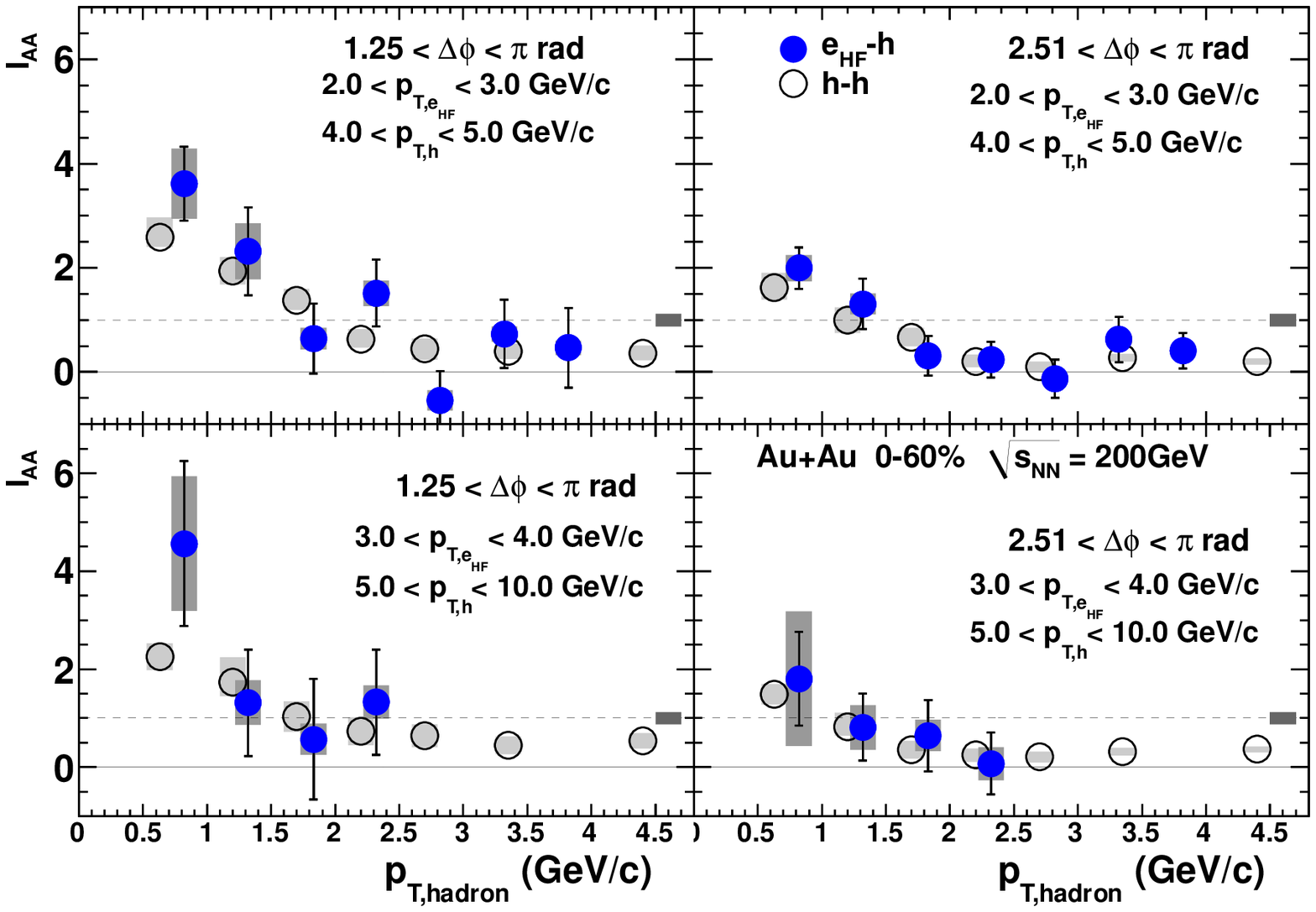}
\caption{\label{away_2panel} (color online) 
$I_{\rm AA}$ as determined from the away-side yields in 
Fig.~\ref{yields_away_2panel}.  Two $\Delta\phi$ ranges are shown: 
1.25$<\Delta\phi<\pi$~rad (left panels) and 2.51$<\Delta\phi<\pi$~rad 
(right panels).  The gray band around unity shows the overall 
normalization uncertainty of 12.4\%, which moves all points together.  
For comparison hadron-hadron $I_{\rm AA}$ values from 
Ref.~\cite{ppg083} are also shown for trigger $p_T$ selections where 
the parent heavy meson has similar $p_T$ to the trigger light hadron 
(see Table~\ref{tabpt}).  Points are slightly shifted horizontally 
for clarity.  The solid horizontal line is at 0 and the dashed 
horizontal line is at 1.}
\end{minipage}
\end{figure*}


\subsection{Au+Au Collisions}

The conditional yields for the near-side (0$<\Delta\phi<$1.25 rad) for 
Au+Au collisions are compared to $p$+$p$ collisions in 
Fig.~\ref{nearyield_auau} for Au+Au collisions with 0--60\% 
centrality.  In addition, to provide a more direct comparison of the 
Au+Au and $p$+$p$ conditional yields we construct the ratio of the 
conditional yield in Au+Au to $p$+$p$:
\begin{equation}
 I_{\rm AA} \equiv \frac{\int Y^{\rm AuAu}_{e_{\rm HF}-h}(\Delta\phi)d\Delta\phi}
 {\int Y^{pp}_{e_{\rm HF}-h}(\Delta\phi)d\Delta\phi}
\end{equation} 
shown in Fig.~\ref{iaa_near}.  In the absence of any nuclear 
effects $I_{\rm AA}$ will be unity.  The near-side $I_{\rm AA}$ is 
consistent with one ($\chi^2/dof$=12.3/7, statistical uncertainties 
only).  Naively, this might be expected since the near-side 
correlations in both $p$+$p$ and Au+Au collisions are expected to 
largely be from the heavy meson decay.  Since the decay length is 
long compared to the size and lifetime of the matter produced in 
Au+Au collisions the subsequent decay of the heavy meson should be 
unmodified by the matter.  However, it is possible that the charm and 
bottom contributions are altered from $p$+$p$ collisions 
due to medium effects (such as different energy loss for charm 
and bottom quarks).  Additionally, the measured hadrons are not solely 
from $D$ and $B$ decay, but also from the fragmentation of the heavy 
quarks and, possibly, from interactions between the heavy quark and 
the matter.  Rather than attempt to disentangle these contributions 
(which would be highly model dependent) we leave it to theoretical 
models to reproduce the $I_{\rm AA}$ with the combined hadron sources.

We compare the $e_{\rm HF}-h$ $I_{\rm AA}$ values to those from 
hadron-hadron collisions at approximately the same meson $p_T$ (see 
Table~\ref{tabpt}).  In hadron-hadron correlations the observed 
$I_{\rm AA}$ has a strong dependence on the $p_T$ of the trigger 
hadron~\cite{ppg083}.  For 2$<p_T<$3~GeV/$c$ electrons the closest 
hadron $p_T$ selection from Ref.~\cite{ppg083} was 4$<p_T<$5~GeV/$c$ 
and for 3$<p_T<$4~GeV/$c$ electrons it was 5$<p_T<$10~GeV/$c$ (see 
Table~\ref{tabpt}).  We observe the near-side $I_{\rm AA}$ for 
heavy-flavor electron triggers to be consistent with those from the 
comparison hadron triggered results, though the present uncertainties 
are too large to be sensitive to the excess seen in the hadron-hadron 
correlations.

\begin{figure}[ht]
\centering
\includegraphics[width=1.0\linewidth]{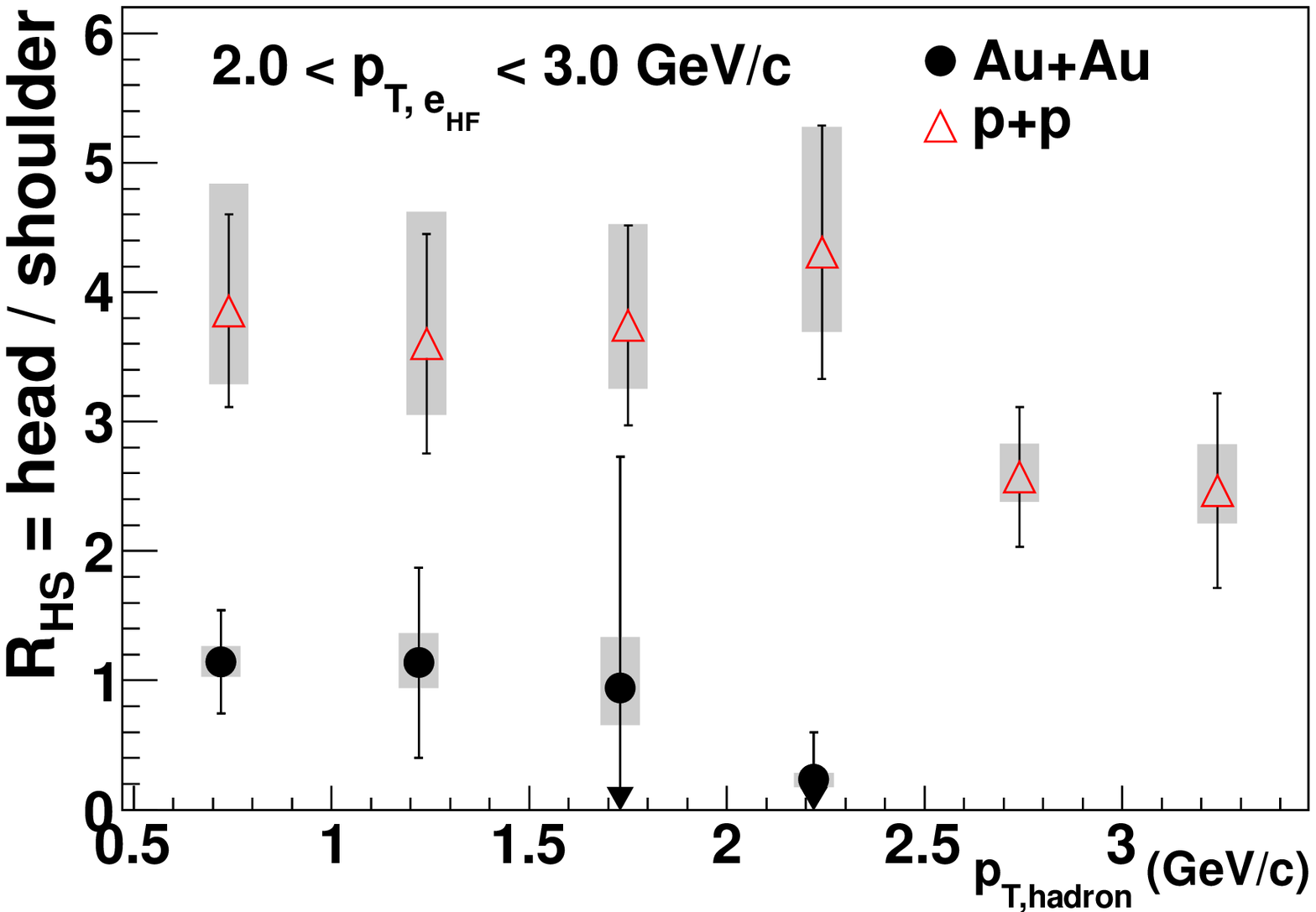}
\caption{(color online) 
Ratio of the yield in the head region per radian to that in 
the shoulder region per radian for Au+Au (black) and $p$+$p$ (red).}
\label{HS_ratio}
\end{figure}

In order to be sensitive to possible modifications of the away-side 
jet shape, we measure the away-side yields in two $\Delta\phi$ ranges 
as shown in Fig.~\ref{yields_away_2panel} for $e_{\rm HF}$ triggers 
with 2.0$<p_{T,e}<$3.0~GeV/$c$ and 3.0$<p_{T,e}<$4.0~GeV/$c$.  The 
wide away-side range, 1.25$<\Delta\phi<\pi$~rad is sensitive to the 
entire modified away-side shape and the smaller away-side range, 
2.51$<\Delta\phi<\pi$~rad, is sensitive to only the $p$+$p$ like part 
of the away-side correlations.  The ratio of conditional yields in 
Au+Au to $p$+$p$ for both away-side $\Delta\phi$ ranges is shown in 
Fig.~\ref{away_2panel}.  $I_{\rm AA}$ is largest at low hadron 
$p_T$ and decreases with increasing hadron $p_T$.  For comparison, 
the hadron-hadron $I_{\rm AA}$ from Ref.~\cite{ppg083} is shown for 
the most closely matched meson $p_T$ selections (see 
Table~\ref{tabpt}).  The heavy-flavor electron triggered $I_{\rm AA}$ 
is consistent with the hadron-hadron $I_{\rm AA}$ when compared at 
similar meson $p_T$ selections.

This similarity could be due to the expected large fraction of gluons 
in the away-side distributions.  If the away-side parton path lengths 
through the matter are similar between heavy-flavor electron and light 
hadron triggers then the corresponding $I_{\rm AA}$ values should also 
be similar.  However, some fraction of the away-side correlations 
should be due to correlated heavy-flavor quarks.  This can be isolated 
in future measurements by triggering on back-to-back heavy-flavor 
electrons.  We conclude that the present measurements are not 
sensitive to any differences caused by back-to-back heavy-flavor 
pairs, either because the differences between away-side heavy-flavor 
electrons and light partons are small or because there are too few of 
them to significantly alter the $I_{\rm AA}$ values.

Motivated by hadron-hadron correlations, we compared the away-side 
jet shape between $p$+$p$ and Au+Au collisions.  To quantify the shape 
differences we construct $R_{HS}$~\cite{ppg083}, which is the 
yield/radian in the head region where the $p$+$p$ jet is peaked (here 
2.51$<\Delta\phi<\pi$ rad) divided by the yield/radian in the 
shoulder region where the enhancement in the Au+Au jet yield is 
observed in hadron-hadron correlations (1.25$<\Delta\phi<$2.51 rad).  
The systematic uncertainties on the ratio are largely correlated 
between the head and shoulder region except for the uncertainty due 
to $v_2$ in Au+Au collisions, which is anti-correlated because of the 
shape of the azimuthal modulation from $v_2$.  In $p$+$p$ collisions 
this ratio is large since the yield in the head region is much larger 
than the yield in the shoulder region.  In hadron-hadron correlations 
for Au+Au collisions $R_{HS}$ is observed to be smaller than in 
$p$+$p$ collisions because of the increased yield in the shoulder 
region~\cite{ppg083}.  Fig.~\ref{HS_ratio} shows $R_{HS}$ for 
$e_{\rm HF}-h$ correlations for 2.0$<p_{T,e}<$3.0~GeV/$c$ as a 
function of the $p_T$ of the associated hadron.  $R_{HS}$ is smaller 
for Au+Au collisions than for $p$+$p$ collisions indicating that a 
similar away-side shape modification takes place for $e_{\rm HF}$ 
triggers as for hadron triggers (the head and shoulder $\Delta\phi$ 
regions are slightly different between this analysis and 
Ref.~\cite{ppg083} preventing a direct comparison).  No $p_{T,h}$ 
dependence of $R_{HS}$ is observed, however the statistical 
uncertainties are quite large.

\section{Conclusion \& Outlook}
\label{conclusions}

Studies of the yields of electrons from the decay of heavy-flavor 
mesons in Au+Au collisions have challenged the picture of 
medium-induced radiative energy loss as the dominant mechanism by 
which high-$p_T$ hadrons are suppressed.  Correlations of hadrons 
from light quark and gluon jets have shown large modifications of the 
correlation patterns between $p$+$p$ and Au+Au collisions.  Studying 
the correlations of electrons from the decay of heavy mesons with 
other hadrons in the event provides more information about how the 
charm and bottom quarks propagate through the matter and how the 
modified correlation structures observed in hadron-hadron 
correlations are produced.  Thus, they are a crucial component of hard 
physics in relativistic heavy-ion collisions.  The interpretation is 
complicated by the ambiguity in the away-side flavor and because the 
electron does not carry all of the parent meson's momentum.  This 
makes understanding $p$+$p$ collisions as a baseline very important.

We have presented first measurements of the azimuthal correlations of 
electrons from heavy-flavor decay with hadrons in both $p$+$p$ and 
Au+Au collisions.  These measurements provide a first step in 
understanding correlations involving open heavy flavor in the hot 
matter produced in heavy-ion collisions.  The Gaussian widths of these 
correlations are consistent with expectations from simulations of the 
fragmentation of heavy quarks and the decay of heavy-flavor mesons, 
except for the lowest $p_T$ electrons and hadrons where some 
differences are observed.  In $p$+$p$ collisions the spectra of 
associated hadrons on both the near and away side are harder than in 
hadron-hadron correlations measured in the same trigger $p_T$ range.  
However, the level of away-side suppression at large $p_T$ is 
consistent between electron and hadron triggers when the trigger 
charged hadron and the parent heavy meson are at approximately the 
same $p_T$.  The ratio of yields in the head region to those in the 
shoulder region decreases from $p$+$p$ to Au+Au collisions in a 
manner qualitatively consistent with hadron-hadron 
collisions~\cite{ppg083}.  Further measurements sensitive to the 
partonic content of the away-side jets (heavy quarks or light quarks 
and gluons) are necessary to determine if this is due primarily to 
cases where the away-side parton is a light quark or gluon or if the 
suppression of away-side heavy-jet fragmentation is similar to those 
of light partons.

Near future measurements of heavy-flavor triggered azimuthal 
correlations hold particular promise.  Data taken in 2010 has 
improved statistics and the HBD was successfully operated allowing 
the rejection of some of the Dalitz and conversion electron 
background.  Additionally, $d$+Au data taken by PHENIX in 2008 will 
help constrain any cold nuclear matter effects.  Such effects are 
expected to be small at midrapidity, but are not well constrained by 
existing data.  In future data taking, the silicon vertex detector 
will be installed, which will enable the separation of electrons from 
$D$ and $B$ decay and increase acceptance for measuring charged 
hadrons.  Application of the techniques developed here on data taken 
with these upgrades in place will allow for more detailed 
heavy-flavor correlation measurements.

\section*{ACKNOWLEDGMENTS}   

We thank the staff of the Collider-Accelerator and Physics
Departments at Brookhaven National Laboratory and the staff of
the other PHENIX participating institutions for their vital
contributions.  We acknowledge support from the 
Office of Nuclear Physics in the
Office of Science of the Department of Energy,
the National Science Foundation, 
a sponsored research grant from Renaissance Technologies LLC, 
Abilene Christian University Research Council, 
Research Foundation of SUNY, 
and Dean of the College of Arts and Sciences, Vanderbilt University 
(USA),
Ministry of Education, Culture, Sports, Science, and Technology
and the Japan Society for the Promotion of Science (Japan),
Conselho Nacional de Desenvolvimento Cient\'{\i}fico e
Tecnol{\'o}gico and Funda\c c{\~a}o de Amparo {\`a} Pesquisa do
Estado de S{\~a}o Paulo (Brazil),
Natural Science Foundation of China (People's Republic of China),
Ministry of Education, Youth and Sports (Czech Republic),
Centre National de la Recherche Scientifique, Commissariat
{\`a} l'{\'E}nergie Atomique, and Institut National de Physique
Nucl{\'e}aire et de Physique des Particules (France),
Ministry of Industry, Science and Tekhnologies,
Bundesministerium f\"ur Bildung und Forschung, Deutscher
Akademischer Austausch Dienst, and Alexander von Humboldt Stiftung (Germany),
Hungarian National Science Fund, OTKA (Hungary), 
Department of Atomic Energy and Department of Science and Technology (India),
Israel Science Foundation (Israel), 
National Research Foundation and WCU program of the 
Ministry Education Science and Technology (Korea),
Ministry of Education and Science, Russia Academy of Sciences,
Federal Agency of Atomic Energy (Russia),
VR and the Wallenberg Foundation (Sweden), 
the U.S. Civilian Research and Development Foundation for the
Independent States of the Former Soviet Union, 
the US-Hungarian Fulbright Foundation for Educational Exchange,
and the US-Israel Binational Science Foundation.

\section*{APPENDIX:  Jet Functions}
\label{jf_appendix}

\begin{figure*}[ht]
\begin{minipage}{1.0\linewidth}
\includegraphics[width=0.6\linewidth]{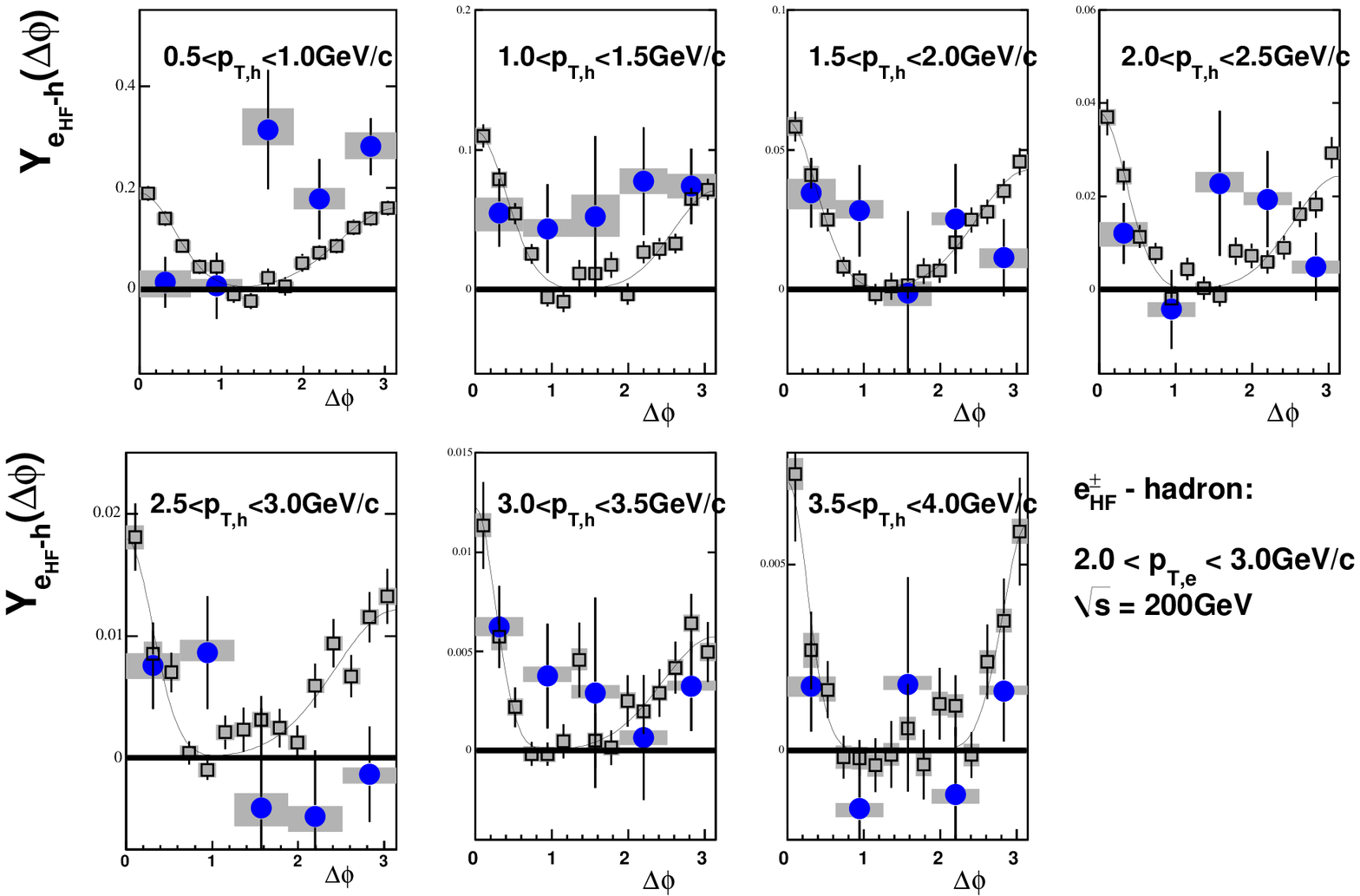}
\caption{(color online) 
$e_{\rm HF}-h$ jet functions for Au+Au (solid blue circles) and $p$+$p$ 
collisions for 2.0--3.0~GeV/$c$
electron triggers and the hadron-$p_T$ bins indicated.}
\label{hf_run7_dphi_rebin_1}
\end{minipage}

\begin{minipage}{1.0\linewidth}
\includegraphics[width=0.6\linewidth]{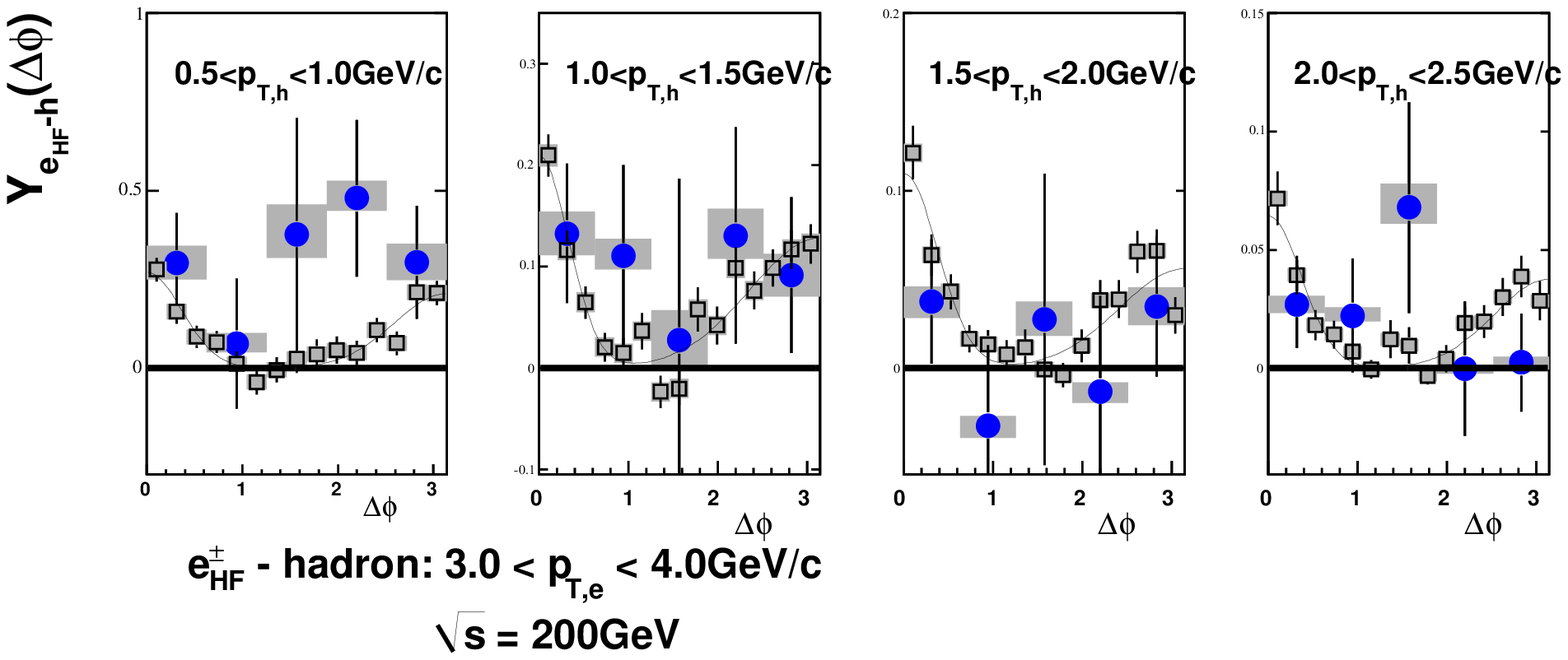}
\caption{(color online)
$e_{\rm HF}-h$ jet functions for Au+Au (solid blue circles) and $p$+$p$ 
collisions for 3.0--4.0~GeV/$c$ electron triggers and the hadron-$p_T$ bins 
indicated.}
\label{hf_run7_dphi_rebin_2}
\end{minipage}

\begin{minipage}{1.0\linewidth}
\includegraphics[width=0.6\linewidth]{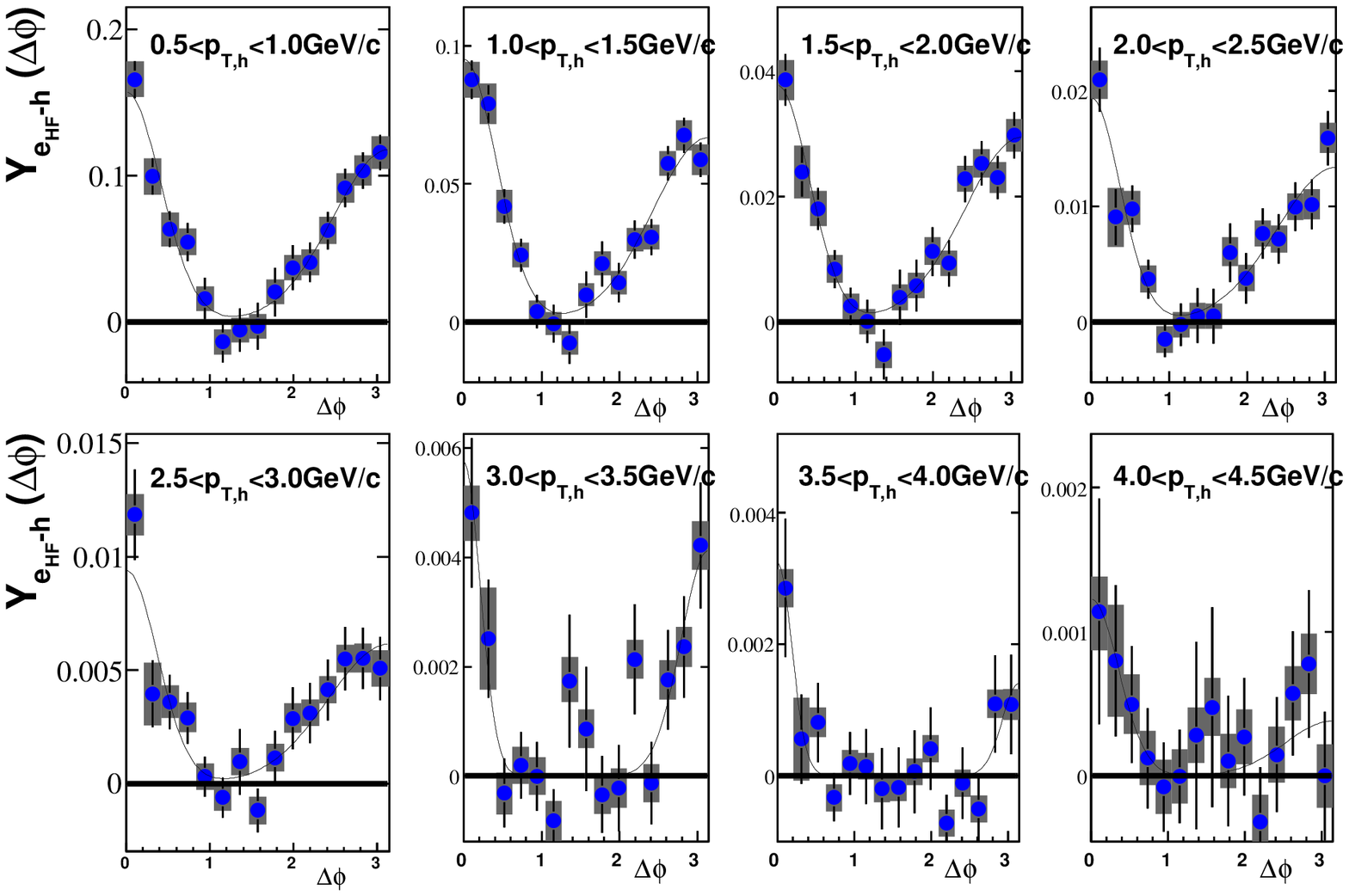}
\caption{\label{hdphi0} (color online)
$e_{\rm HF}-h$ jet functions in $p$+$p$ collisions for 1.5--2.0~GeV/$c$ 
electron triggers and the hadron-$p_T$ bins indicated.}
\end{minipage}
\end{figure*}

\begin{figure*}[ht]
\begin{minipage}{1.0\linewidth}
\includegraphics[width=0.6\linewidth]{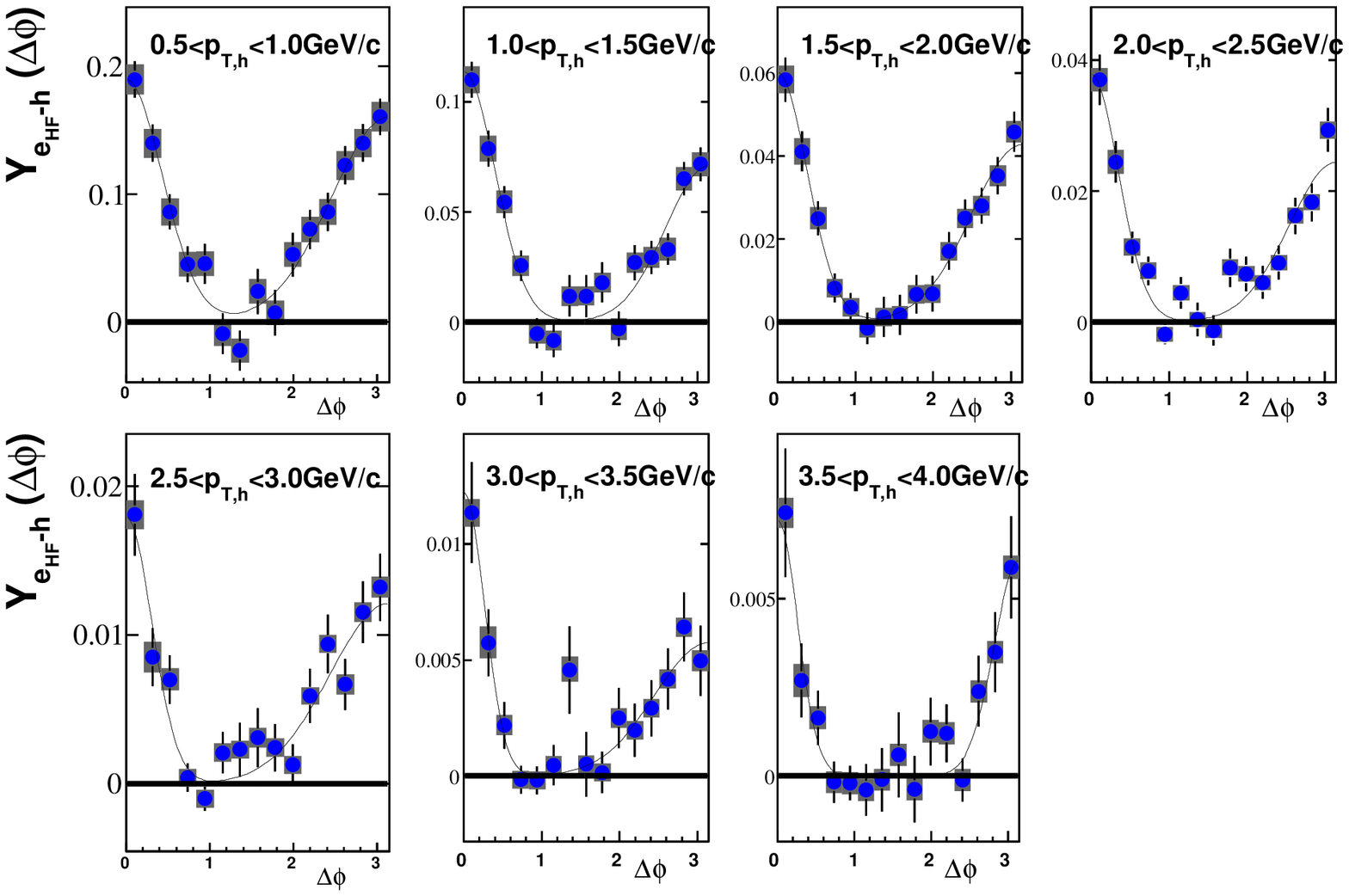}
\caption{\label{hdphi1} (color online)
$e_{\rm HF}-h$ jet functions in $p$+$p$ collisions for 2.0--3.0~GeV/$c$ 
electron triggers and the hadron-$p_T$ bins indicated.}
\end{minipage}

\begin{minipage}{1.0\linewidth}
\includegraphics[width=0.6\linewidth]{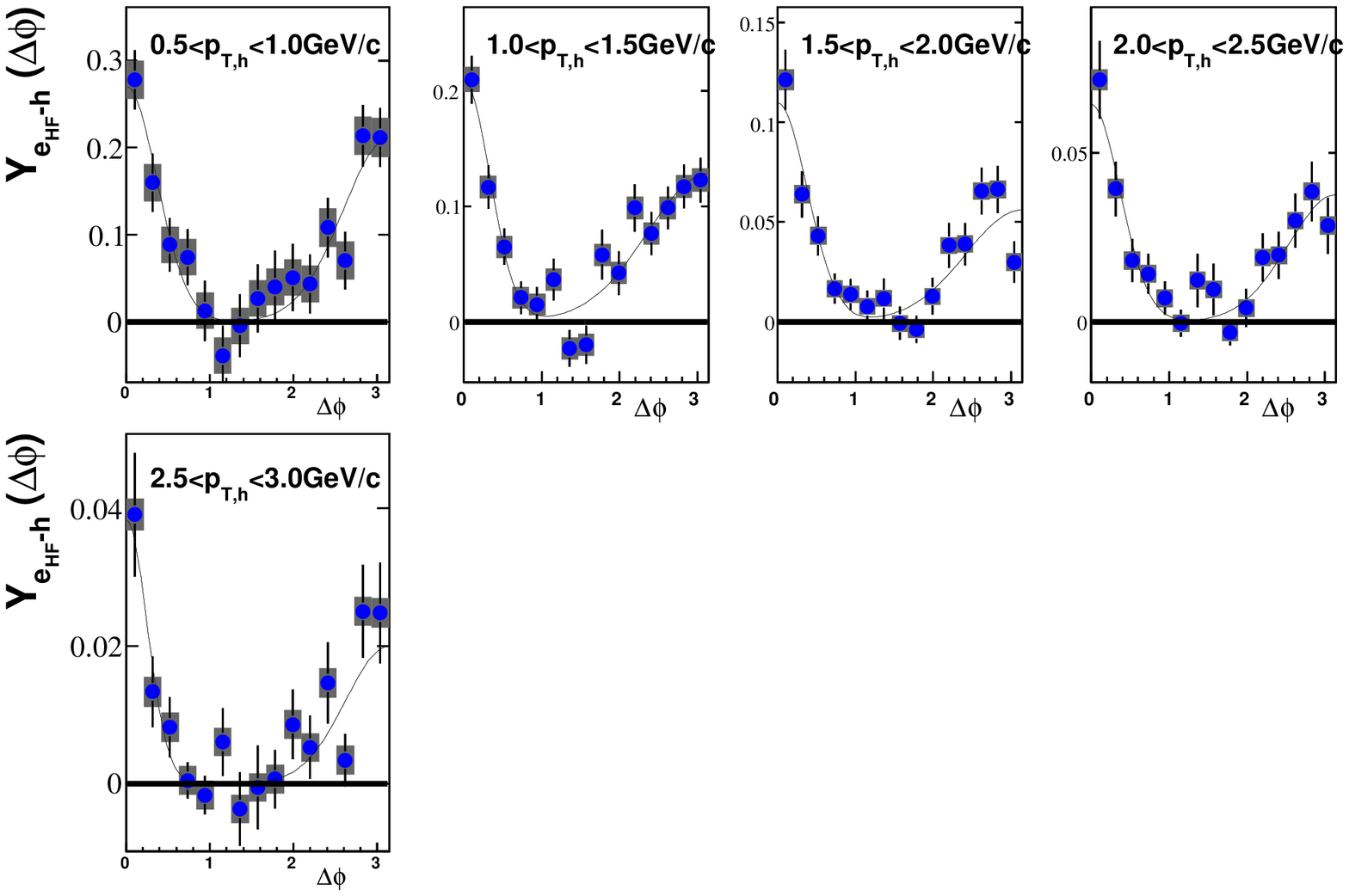}
\caption{\label{hdphi2} (color online)
$e_{\rm HF}-h$ jet functions in $p$+$p$ collisions for 3.0--4.0~GeV/$c$ 
electron triggers and the hadron-$p_T$ bins indicated.}
\end{minipage}

\begin{minipage}{1.0\linewidth}
\includegraphics[width=0.6\linewidth]{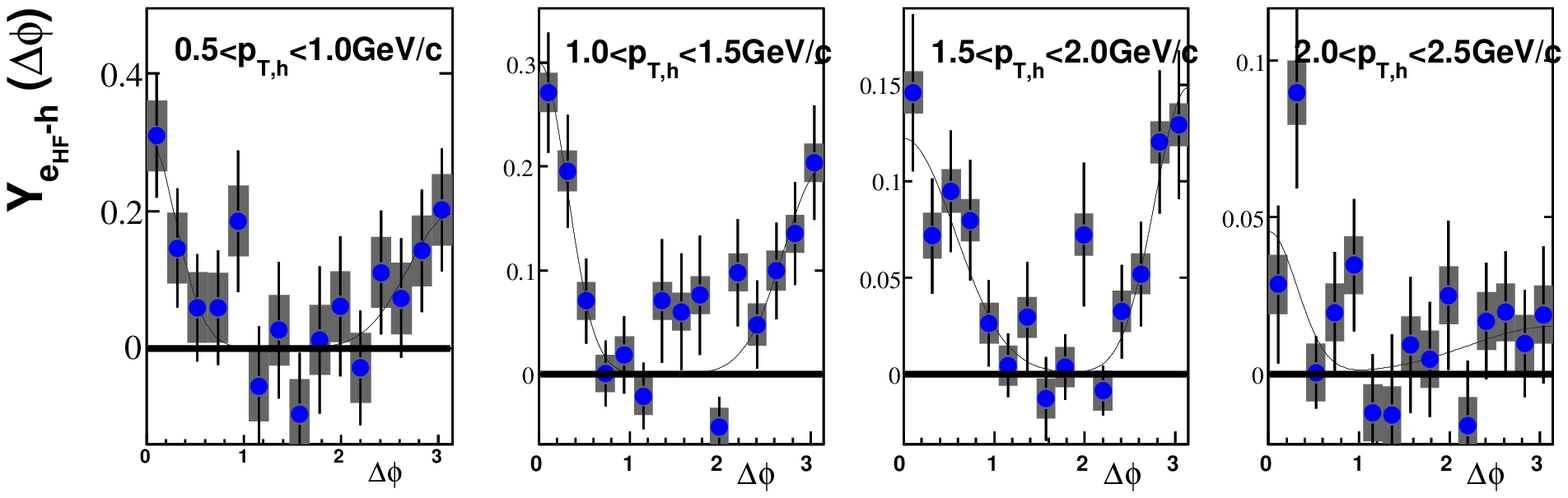}
\caption{\label{hdphi3} (color online)
$e_{\rm HF}-h$ jet functions in $p$+$p$ collisions for 4.0--4.5~GeV/$c$ 
electron triggers and the hadron-$p_T$ bins indicated.}
\end{minipage}
\end{figure*}

Figures~\ref{hf_run7_dphi_rebin_1} and \ref{hf_run7_dphi_rebin_2} show 
comparisons of $e_{\rm HF}-h$ jet functions for Au+Au and $p$+$p$ collisions for 
the indicated electron triggers and hadron-$p_T$ bins. 
Figures~\ref{hdphi0}--\ref{hdphi3}
show the $e_{\rm HF}-h$ jet 
functions for $p$+$p$ collisions only for the indicated electron triggers and 
hadron-$p_T$ bins.


\clearpage %


\end{document}